\documentclass[11pt]{article}
\pdfoutput=1
\usepackage[utf8]{inputenc}
\usepackage{multirow}
\usepackage{amsmath,amsfonts,amssymb}
\usepackage{array}
    \newcolumntype{C}[1]{>{\centering\arraybackslash$}p{#1}<{$}}
\usepackage{nicematrix}
\usepackage{float}
\usepackage{comment}
\usepackage{graphicx}
\usepackage{cite}
\usepackage{pdflscape}
\usepackage{psfrag}
\usepackage{amsthm}
\usepackage{enumerate}
\usepackage{arydshln}
\usepackage{pifont}
    
\usepackage{soul}
\usepackage{slashed}
\usepackage{subcaption}
\usepackage{mathrsfs}
\usepackage{ytableau}
 \usepackage{a4wide}
  \usepackage{tikz}
  \usepackage{tikz-cd}
  \usetikzlibrary{shapes.geometric}
  \usepackage{tcolorbox}
  \usepackage[table]{xcolor}
  \usepackage{colortbl}
  \definecolor{dark-gray}{gray}{0.20}
  \definecolor{gray}{gray}{0.30}
  \definecolor{light-gray}{gray}{0.80}
  \definecolor{dark-red}{rgb}{0.7,0,0}
  \definecolor{dark-green}{rgb}{0.1,0.4,0}
  \definecolor{dark-blue}{rgb}{0.3,0.3,0.7}
  \definecolor{light-blue}{rgb}{0.8,0.8,1}
      \definecolor{swamp}{RGB}{240, 199, 197}

\usepackage[framemethod=TikZ]{mdframed}
\mdfdefinestyle{function}{%
    nobreak=true,
    skipabove=1em,
    linecolor=black!80!white,
    outerlinewidth=0.5pt,
    roundcorner=1pt,
    innertopmargin=10pt,
    innerbottommargin=10pt,
    innerrightmargin=10pt,
    innerleftmargin=10pt,
    backgroundcolor=yellow!10!white}
\mdfdefinestyle{code}{%
    nobreak=true,
    skipabove=1em,
    linecolor=black,
    outerlinewidth=0.25pt,
    roundcorner=0pt,
    innertopmargin=10pt,
    innerbottommargin=10pt,
    innerrightmargin=10pt,
    innerleftmargin=10pt,
    backgroundcolor=gray!2!white}

\newcommand{\function}[1]{\textcolor{dark-blue}{\texttt{#1}}}
\usepackage{mmacells}
\mmaDefineMathReplacement[≤]{<=}{\leq}
\mmaDefineMathReplacement[≥]{>=}{\geq}
\mmaDefineMathReplacement[≠]{!=}{\neq}
\mmaDefineMathReplacement[→]{->}{\to}[2]
\mmaDefineMathReplacement[⧴]{:>}{:\hspace{-.2em}\to}[2]
\mmaDefineMathReplacement{∉}{\notin}
\mmaDefineMathReplacement{∞}{\infty}
\mmaDefineMathReplacement{��}{\mathbbm{d}}
\mmaSet{
  morefv={gobble=2},
  linklocaluri=mma/symbol/definition:#1,
  morecellgraphics={yoffset=1.9ex}
}
 
\usepackage{pifont}

\usepackage{newunicodechar} 
\usepackage{setspace}

\usepackage{ifthen}
\usepackage{longtable}

\newcommand{\be}{\begin{equation}}
\newcommand{\ee}{\end{equation}}
\newcommand{\eq}[1]{(\ref{#1})}

\def\be{\begin{equation}}
\def\ee{\end{equation}}
\def\bea{\begin{eqnarray}}
\def\eea{\end{eqnarray}}

\newcommand{\shortexact}[3]{
	1 \longrightarrow #1
	\longrightarrow  #2
	\longrightarrow #3
	\longrightarrow 1
}

\newcommand{\D}[1][\gamma]{\mathbf{D}_{\bf #1}}
\newcommand{\bvec}[1][\gamma]{\vec{b}_{\bf #1}}
\newcommand{\RFM}[2][T]{%
  \ifthenelse{\equal{#1}{T}}%
    {T^{#2}/\Gamma}%
    {\frac{\mathbb{R}^{#2}}{\mathcal{B}}}%
}
\newcommand{\Tr}[2]{{\rm Tr_{\bf #1}}(#2)}

\newcommand{\Vcas}{V_{\text{Cas}}}

\newcommand{\R}{\mathbb{R}}

\newcommand{\Z}{\mathbb{Z}}

\newcommand{\Spin}{\mathit{Spin}}

\newcommand{\F}[1]{\mathcal{F}_{#1}}

\numberwithin{equation}{section}

\usepackage{jheppub}

\usepackage{cleveref}

\hypersetup{
	colorlinks=true,
	linkcolor=dark-blue,
	citecolor=dark-red,
	urlcolor=dark-green,
	linktoc=page,
}

\crefname{appendix}{Appendix}{Appendices}

\hypersetup{
    pdftitle={CasimirRFM: A Mathematica package for Riemann-flat compactifications with Casimir energies}, 
    pdfauthor={Bruno Valeixo Bento}, 
    pdfsubject={Mathematica package documentation}, 
    pdfkeywords={CasimirRFM, Casimir energy, compactification, flat manifolds, Mathematica}
}

\title{\texttt{CasimirRFM}: A Mathematica package for Riemann-flat compactifications with Casimir energies}

\author{Bruno Valeixo Bento} 
\affiliation{Instituto de F\'{i}sica Te\'{o}rica IFT-UAM/CSIC,
C/ Nicol\'{a}s Cabrera 13-15, Campus de Cantoblanco, 28049 Madrid, Spain}

\emailAdd{bruno.bento@ift.csic.es}

\abstract{CasimirRFM is a Mathematica package for the study of compactifications on Riemann-flat manifolds and the computation of one-loop Casimir energies in higher-dimensional field theories and supergravity. It implements an efficient numerical evaluation of lattice sums using Ewald summation, and computes both lower-dimensional Casimir potentials and local higher-dimensional Casimir energy densities, allowing for general massless spectra and twisted boundary conditions. The package provides tools for constructing and analysing the finite groups defining these manifolds, determining invariant metrics and cohomology bases, identifying compatible spin structures, and computing moduli-space metrics. Moreover, it evaluates the traces of holonomy elements in the graviton, $p$-form, spinor, and Rarita-Schwinger representations. We describe the mathematical formulation and numerical implementation of the package, and illustrate its use through a compactification of Type IIB supergravity on $T^6/\mathbb{Z}_8$, including the computation of the Casimir potential and the visualization of localised Casimir-brane contributions.}

\begin{document}
\emergencystretch 3em
\hypersetup{pageanchor=false}
\makeatletter
\let\old@fpheader\@fpheader
\preprint{IFT-26-99
}

\makeatother

\maketitle

\section{Introduction \& Quick description}
\label{sec:introduction}

\texttt{CasimirRFM} is a Mathematica package built for the computation of Casimir energies in Riemann-flat compactifications of $D$-dimensional supergravity theories. It implements the methods developed in \cite{ValeixoBento:2025yhz}, with particular emphasis on the one-loop Casimir contribution from the massless spectrum of higher-dimensional supersymmetric field theories. The package includes a broad set of functions that can be split in three groups.
\begin{enumerate}
    \item \textbf{Functions for Riemann-flat manifolds}

    These functions can be used to study Riemann-flat manifolds and their properties. Some of these focus on the holonomy groups defining the Riemann-flat manifolds (RFMs), which apply more generally to finite groups $\Gamma$, and are listed in Table~\ref{tab:RFM-group-functions}; others refer to properties of the manifold itself, and are listed separately in Table~\ref{tab:RFM-compactification-functions}.

    \item \textbf{Functions for Casimir potentials}

    These functions compute the Casimir potentials and energy densities arising in compactifications on RFMs. They are based on an efficient implementation of Ewald's summation method \cite{ewaldOG,NIJBOER1957309}, and are listed in Table~\ref{tab:Casimir-potential-functions}.

    \item \textbf{Functions for Lorentz group traces}

    The computation of Casimir potentials requires taking the trace of group elements $\gamma\in\Gamma\subset SO(k)$ in different representations belonging to the massless spectrum of the higher-dimensional theory. These functions compute the traces in the graviton, $p$-form, spinor and Rarita-Schwinger representations, and are listed in Table~\ref{tb:list-functions-traces}.
\end{enumerate}

The computation of Casimir energies can be made more efficient by combining the functions implemented in the package with Mathematica's parallelization functions; it is useful to run multiple evaluations in parallel using Mathematica functions such as 
\begin{itemize}
    \item[] \texttt{Parallelize, ParallelEvaluate, ParallelTable, ParallelSum}\,.
\end{itemize}
For maximum flexibility, \texttt{Parallelize} is not used internally in any of the function definitions, apart from \hyperlink{RFMGrad}{\function{RFMGrad}} and \hyperlink{RFMHessian}{\function{RFMHessian}}. 
It is helpful to start by launching all kernels and loading the package in each of them,

\begin{mmaCell}[
  moredefined={f,LaunchKernels,ParallelEvaluate,DistributedContexts,CasimirRFM,PackageDirectory},
  excessargument=n,
]{Input}
  LaunchKernels[];
  ParallelEvaluate[<< CasimirRFM\`\,];
  DistributedContexts = None;
  empty
\end{mmaCell}

\subsection*{Installation}

To install the package 
\begin{enumerate}
    \item Download \href{https://github.com/bruno-valeixo-bento/CasimirRFM}{\texttt{CasimirRFM.wl}};
    \item Copy \texttt{CasimirRFM.wl} to \texttt{\$UserBaseDirectory/Applications/CasimirRFM/} 
\end{enumerate} 
It can then be loaded in any \texttt{Mathematica} notebook with the usual command\footnote{The symbol at the end of the package name used by Mathematica is \texttt{U+0060 Grave Accent}.}
\begin{mmaCell}{Input}
  \mmaDef{<< CasimirRFM\` }
\end{mmaCell}
The package includes a description and template for every function, which can be accessed within the notebook as
{
\flushleft\includegraphics[width=0.6\textwidth]{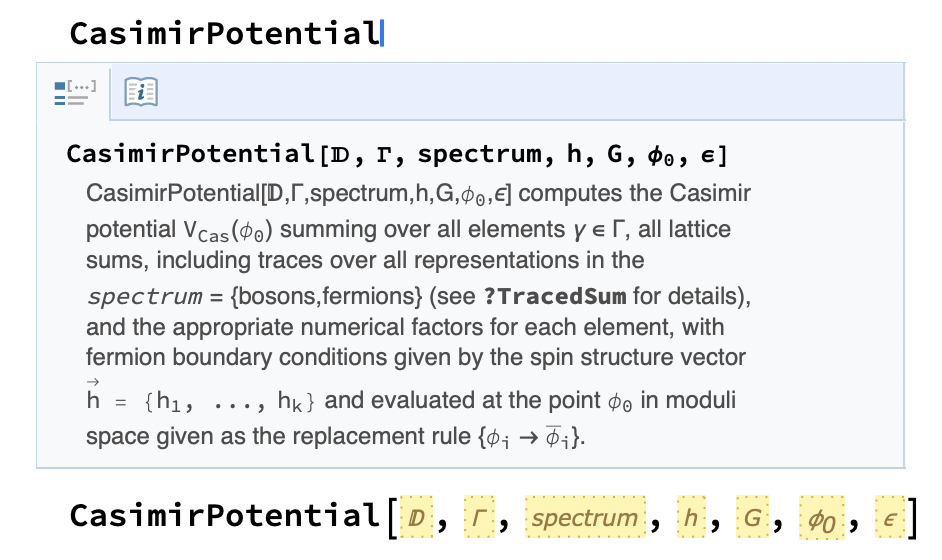}
\vspace{1em}}

In the following sections we will describe each function in more detail, including its input, output, options, and mathematical background behind its implementation. For easier reference, we provide as supplementary material the notebook \texttt{CasimirRFM-Template.nb}, listing all functions, their descriptions and concrete examples of their use. We also include a notebook \texttt{CasimirRFM-TypeIIB-Example.nb} with the example provided in Section~\ref{sec:example-TypeIIB}.

\pagebreak

\begin{table}[H]
    \renewcommand*{\arraystretch}{2}
    \centering
    \begin{tabular}{lp{9cm}} 
        \hline
        \textbf{Function} & \textbf{Returns} \\ \hline 
        \hyperlink{DefGenerator}{\function{DefGenerator[M,$\sigma_\gamma$]}} & Generator \texttt{g = \{$\D$,\{$b_1,\ldots,b_k$\},$\sigma_\gamma$\}} built from the matrix \texttt{M} = {\renewcommand*{\arraystretch}{1} $\begin{pmatrix} \D & \bvec \\  0 & 1 \end{pmatrix}$} and the choice of spin lift $\sigma_\gamma\in\{0,\frac12\}$ (set to 0 by default). \\ \hline
         \hyperlink{ActsFreelyQ}{\function{ActsFreelyQ[$\Gamma$]}} & \texttt{True} if every non-identity element of $\Gamma$ acts without fixed points, and \texttt{False} otherwise. \\ \hline
         \hyperlink{ElementAction}{\function{ElementAction[g][z]}} &  Vector \texttt{g(z)} obtained by acting with element \texttt{g} on the vector \texttt{z}. \\ \hline
         \hyperlink{ComposeElements}{\function{ComposeElements[g$_1$,g$_2$]}} & Element $\texttt{g}_1\circ \texttt{g}_2$.
         \\ \hline
         \hyperlink{ElementOrder}{\function{ElementOrder[g]}} & Order $k$ of \texttt{g}, such that $\texttt{g}^k=\textbf{1}$, as an affine torus isometry, not taking into account the spin lift.  \\ \hline
         \hyperlink{GetAllowedShifts}{\function{GetAllowedShifts[D$_\texttt{g}$]}} & Set of shift vectors $\bvec[g]$ satisfying $\D[g]\,\bvec[g]=\bvec[g]\,(\text{mod}\,\Lambda_{\perp})$ for which the resulting space-group action is fixed-point free. \\ \hline
         \hyperlink{GetGroup}{\function{GetGroup[\{g$_1$,g$_2$,\ldots\}]}} & Finite group generated by \texttt{\{g$_1$,g$_2$,\ldots\}}. \\ \hline
         \hyperlink{NormalizerQ}{\function{NormalizerQ[$\Gamma$,\{Q,q\}]}} & \texttt{True} if the transformation $f(\vec{z})=Q\,\vec{z}+\vec{q}$ represented by \texttt{\{Q,q\}} is in the normalizer of group $\Gamma$; \texttt{`True for point group only'} if $f(\vec{z})$ normalizes the point group, but not the space group; and \texttt{False} otherwise. \\ \hline
         \hyperlink{GetCompanionMatrix}{\function{GetCompanionMatrix[p(x),x]}} & Companion matrix of the polynomial \texttt{p(x)} in the form of a \texttt{SparseArray}. \\ \hline
         \hyperlink{FiniteOrderMatrices}{\function{FiniteOrderMatrices[k]}} & Block-diagonal representatives of the rational conjugacy classes of finite-order $k\times k$ integer matrices, constructed as direct sums of companion matrices of cyclotomic polynomials. The output is a list of pairs \texttt{\{order,matrix\}}. \\ \hline
    \end{tabular}
    \caption{Group functions for Riemann-flat manifolds, including input arguments, defaults and output. These functions focus on the holonomy groups defining the RFMs, and apply more generally to the study of finite groups $\Gamma$. Detailed descriptions of these functions and their implementation can be found in Sections \ref{sec:RFMs} and \ref{sec:group-functions}.}
    \label{tab:RFM-group-functions}
\end{table}

\begin{table}[H]
    \renewcommand*{\arraystretch}{2}
    \centering
    \begin{tabular}{lp{8cm}} 
        \hline
        \textbf{Function} & \textbf{Returns} \\ \hline 
         \hyperlink{CasimirBranes}{\function{CasimirBranes[g]}} & Coordinates of Casimir branes of element \texttt{g}. \\ \hline
         \hyperlink{GetInvariantMetric}{\function{GetInvariantMetric[\{g$_1$,g$_2$,\ldots\}]}} & \texttt{\{number of moduli,\{c$_1$,c$_2$,\ldots\},G\}} where \texttt{G} is the most general metric invariant under the group generated by \texttt{\{g$_1$,g$_2$,\ldots\}}, and parameterised by the geometric moduli \texttt{\{c$_1$,c$_2$,\ldots\}}. \\ \hline
         \hyperlink{GetModuliModuliSpaceMetric}{\function{GetModuliSpaceMetric[D,G,\{c$_i$\},print]}} & Metric on the moduli space parameterised by \{c$_i$\}, upon compactifying a \texttt{D}-dimensional theory on an RFM with metric \texttt{G}. If \texttt{print = True}, the kinetic terms are printed (\texttt{print = False} by default). \\ \hline
         \hyperlink{GetInvariantForms}{\function{GetInvariantForms[\{g$_1$,g$_2$,\ldots\}]}} & \texttt{\{Betti numbers, \{basis of $H^0(\F{k},\R)$, \ldots , basis of $H^k(\F{k},\R)$\}, 
         \texttt{basis form coefficients}\}}, providing a basis of $p$-forms invariant under the action of the group with generators \texttt{\{g$_1$,g$_2$,\ldots\}}. \\ \hline
         \hyperlink{SpinStructures}{\function{SpinStructures[\{g$_1$,g$_2$,\ldots\}]}} & Association $$\langle|\texttt{"GeneralForm"}\to\texttt{h}\,,\texttt{"Allowed"}\to \texttt{list}|\rangle$$
          of spin structures/boundary conditions on $T^k$ compatible with the quotient $T^k/\Gamma$, with $\Gamma$ generated by \texttt{\{g$_1$,g$_2$,\ldots\}}. The \texttt{"GeneralForm"} key points to the most general structure \texttt{\{h$_1$,\ldots,h$_k$\}}; the \texttt{"Allowed"} key points to a list of all allowed spin structures. \\ \hline
    \end{tabular}
    \caption{Compactification functions for Riemann-flat manifolds, including input arguments, defaults and output. These functions focus on the compactification of a $D$-dimensional field theory on a $k$-dimensional RFM. Detailed descriptions of these functions and their implementation can be found in Section \ref{sec:compactification-functions}.}
    \label{tab:RFM-compactification-functions}
\end{table}

\begin{table}[H]
    \renewcommand*{\arraystretch}{1.5}
    \centering
    \begin{tabular}{p{6.5cm}p{8cm}} 
        \hline
        \textbf{Function} & \textbf{Returns} \\ \hline 
         \hyperlink{Ewald}{\function{Ewald[c,h,s,$\alpha$,G,$\varepsilon$]}} &  Lattice sum $\sum_{\vec{n}\in\Z^k} \frac{e^{2\pi i \vec{h}\cdot\vec{n}}}{|\vec{n}+\vec{c}|^{2s}_{\texttt{G}}}$ using numerical implementation of Ewald's summation method. The parameter $\alpha$ controls the split between real and reciprocal space contributions in Ewald summation. \\ \hline
         \hyperlink{ReducedLatticeSum}{\function{ReducedLatticeSum[D,g,h,G,$\varepsilon$]}} & Integrated lattice sum $$\int_{[0,1]^k}\,d^kz\,\sum_{\vec{n}\in\Z^k} \frac{e^{2\pi i \vec{h}\cdot\vec{n}}}{|(\mathbf{1}-\D[g])\vec{z}+\vec{n}-\bvec[g]|^{D}_{\texttt{G}}}$$ projecting onto the \texttt{g}-invariant lattice and collapsing the sum to the lowest dimension possible, before summing numerically using \hyperlink{Ewald}{\function{Ewald}}.\\ \hline
         \hyperlink{TracedSum}{\function{TracedSum[\phantom{D,g,spectrum,hspin,G, $\phi_0$,} \phantom{Tra}D,g,spectrum,hspin,G,$\phi_0$,$\varepsilon$]}} & Sum of lattice sum contributions (summed numerically using \hyperlink{ReducedLatticeSum}{\function{ReducedLatticeSum}}) over all representations in \texttt{spectrum} with the appropriate traces of \texttt{g} and boundary conditions; \texttt{hspin} encodes the RFM spin structure and sets the default boundary conditions for fermions; $\phi_0$ is the point in moduli space where the symbolic metric \texttt{G} is evaluated, given as a replacement rule for the moduli. \\ \hline
         \hyperlink{CasimirPotential}{\function{CasimirPotential[\phantom{D,$\Gamma$,spectrum,} \phantom{Cas}D,$\Gamma$,spectrum,hspin,G,$\phi_0$,$\varepsilon$]}} & Casimir potential $\Vcas$ given by the sum of lattice sums over all elements $\gamma\in\Gamma$, each computed using \hyperlink{TracedSum}{\function{TracedSum}}. \\ \hline
         \hyperlink{CasimirEnergyDensity}{\function{CasimirEnergyDensity[\phantom{D,$\Gamma$,spectrum,} \phantom{Cas}D,$\Gamma$,spectrum,hspin,G,$\phi_0$,z,$\varepsilon$]}} & Casimir energy density $\rho_{\rm Cas}(\vec{z})$ at a point $\vec{z}\in T^k/\Gamma$ given by the sum  of lattice sums over all elements $\gamma\in\Gamma$. \\ \hline
         \function{RFMGrad[V,$\vec{x}_0$,$\delta \vec{x}$]} & Gradient vector $\vec{\partial}\,V|_{x_0}$ normalized by $V$ computed numerically at $\vec{x}_0$, using values of $V$ a distance $\delta \vec{x}$ away from $\vec{x}_0$ (default $\delta\vec{x}=10^{-4}\cdot\vec{1}$). \\ \hline
         \function{RFMHessian[V,$\vec{x}_0$,$\delta \vec{x}$]} & Hessian matrix $\partial_i\partial_j V|_{x_0}$ computed numerically at $\vec{x}_0$, using values of $V$ a distance $\delta\vec{x}$ away from $\vec{x}_0$ (default $\delta\vec{x}=10^{-4}\cdot\vec{1}$). \\ \hline
    \end{tabular}
    \caption{Functions for the computation of lattice sums and Casimir potentials, including input arguments, defaults and output.
    The parameter $\varepsilon$ controls the estimated truncation of the numerical sums such that $\varepsilon = 10^{-p}$ targets roughly $p$ decimal digits ($\varepsilon=10^{-4}$ by default). Detailed descriptions of these functions and their implementation can be found in Section \ref{sec:Casimir}.}
    \label{tab:Casimir-potential-functions}
\end{table}

\begin{table}[H]
    \renewcommand*{\arraystretch}{1.5}
    \centering
    \begin{tabular}{lp{9cm}} 
        \hline
        \textbf{Function} & \textbf{Returns} \\ \hline 
         \hyperlink{TraceSym}{\function{TraceSym[g,D]}} & Trace of element $\texttt{g}\in\Gamma$ in the symmetric traceless (graviton) representation of $SO(D-2)$. \\ \hline
         \hyperlink{TraceForm}{\function{TraceForm[p,g,D]}} & Trace of element $\texttt{g}\in\Gamma$ in the antisymmetric $p$-form representation of $SO(D-2)$. \\  \hline
         \hyperlink{TraceSpinor}{\function{TraceSpinor[g,D,Weyl]}} & Trace of element $\texttt{g}\in\Gamma$ in the spinor representation of $SO(D-2)$, taking into account the choice of spin lift $\pm1$ of \texttt{g}; set parameter \texttt{Weyl = True} for a Weyl spinor (\texttt{Weyl = False} by default). \\ \hline
         \hyperlink{TraceRS}{\function{TraceRS[g,D,Weyl]}} & Trace of element $\texttt{g}\in\Gamma$ in the Rarita-Schwinger (gravitino) representation of $SO(D-2)$, taking into account the choice of spin lift $\pm1$ of \texttt{g}; set parameter \texttt{Weyl = True} for a Weyl spinor (\texttt{Weyl = False} by default). \\ \hline
         \hyperlink{TraceBosons}{\function{TraceBosons[bosons,g,D]}} & Sum of traces of element $\texttt{g}\in\Gamma$ under the bosonic representations listed in \texttt{bosons} in $D$ dimensions; \texttt{bosons} accepts association-based field specifications or the list shorthand notation described in Section \ref{sec:traces}. \\ \hline
         \hyperlink{TraceFermions}{\function{TraceFermions[fermions,g,D]}} & Sum of traces of the (spin lift of) element $\texttt{g}\in\Gamma$ under the fermionic representations listed in \texttt{fermions} in $D$ dimensions; \texttt{fermions} accepts association-based field specifications or the list shorthand notation described in Section \ref{sec:traces} \\ \hline
    \end{tabular}
    \caption{Functions for the computation of Lorentz group traces under different representations, including input arguments, defaults and output. Detailed descriptions of these functions and their implementation can be found in Section \ref{sec:traces}.}
    \label{tb:list-functions-traces}
\end{table}

\section{Detailed description of package functions}
\label{sec:details}

In this section we will give a detailed description of the functions implemented in the \texttt{CasimirRFM} package, and the background---regarding finite groups, RFMs, spin structures, traces of elements of $SO(D-2)$ under different representations, the Ewald summation method and other connected topics---necessary to understand both how to use it and how it is built. 
The detailed descriptions are intended both to support use of the current package and to facilitate future extensions.

\subsection{Riemann-flat manifolds (RFMs) and finite groups}
\label{sec:RFMs}

Every Riemann-flat manifold (RFM) can be defined as the quotient of a torus $T^k$ by a \emph{finite} group $\Gamma$ of isometries of $T^k$, i.e. isometries of $\R^k$ that preserve the torus lattice $\Lambda$ such that $T^k=\R^k/\Lambda$ \cite{charlap1,charlap2,borwein2013lattice,ValeixoBento:2025yhz},
\begin{equation}
    \mathcal{F}_k=T^k/\Gamma \,.
    \label{rfmdef}
\end{equation} 
The group $\Gamma$---also called the point group or holonomy group of the RFM---acts freely on the torus, meaning that it has no fixed points; the quotient is a smooth manifold rather than a singular orbifold. 
An element $\gamma\in\Gamma$ is represented by the pair $(\D,\bvec)$ corresponding to the affine transformation 
\begin{equation}
    \vec{z}\to \iota_\gamma(\vec{z})=\D\,\vec{z}+\bvec
    \label{eq:affine-tranform}
\end{equation}
of points in $T^k$, where $\bvec$ is taken modulo $\Lambda$-translations and $\D$ is orthogonal. We will focus on orientable RFMs, for which $\D\in SO(k)$.

It is always possible to perform a change of coordinates in the ambient space $\mathbb{R}^k$ such that the lattice $\Lambda$ becomes exactly $\mathbb{Z}^k$, the standard integer lattice, at the price of having an ambient metric and corresponding induced metric on the RFM that are no longer standard, but rather some flat metric $G$ that must be determined. In this basis, the matrices $\D\in SL(k,\mathbb{Z})$ preserve the lattice $\mathbb{Z}^k$; from the representation theory of conjugacy classes of such matrices, we know that they are all conjugate to the block-diagonal form \cite{laffeylectures}
\begin{equation} 
    \mathbf{D}_\gamma\sim \begin{pNiceArray}{ccc|ccc}[columns-width=1.5em]
    \mathbf{B}_1& 0 & 0 & \vdots  & \vdots & \vdots \\
    0 & \ddots & 0 & \vec{v}_1 & \cdots & \vec{v}_l \\
    0 & 0 & \mathbf{B}_i & \vdots  & \vdots & \vdots \\ \hline
    0 & \cdots & 0 & \ddots&\vdots &  \\
    \vdots & 0 & \vdots & \cdots & \mathbf{I}_{l\times l} & \cdots \\
    0 &\cdots&0& &\vdots & \ddots\end{pNiceArray} \,,
    \label{eq:D-canonical-form}
\end{equation}
where the diagonal blocks $\mathbf{B}_i$ are standard and determined by the order of the matrix \cite{laffeylectures}, and $l$ is the dimension of its invariant subspace. The only ambiguity is the choice of the upper-right entries, which must be analysed on a case-by-case basis \cite{ValeixoBento:2025yhz}. Note that there may not exist a similarity transformation that simultaneously puts  all $\mathbf{D}_\gamma$ in the form \eq{eq:D-canonical-form} when $\Gamma$ is non-abelian. In all that follows and in the code implementation we work in the basis with $\Lambda=\Z^k$, which is the torus lattice over which we perform the lattice sums. 

When working with spinors, we must also consider the spin lift of $\Gamma$. 
Spinor representations are representations of $Spin(n)$, the double cover of $SO(n)$, rather than ordinary representations of $SO(n)$. 
The group $\Spin(n)$ is such that there exists a short exact sequence of groups
\begin{align*}
    \shortexact{\Z_2}{Spin(n)}{SO(n)} \,,
\end{align*}
so that the spin lift of $\Gamma$ is a $\Z_2$ extension of $\Gamma$, with a transformation $\D$ having two spin lifts $\pm\mathcal{D}_\gamma$ related by a sign (see \cite{lutowski2015spin,ValeixoBento:2025yhz} and Appendix B of \cite{Fraiman:2026irf} for more details). A consistent lift must respect all group relations \cite{lutowski2015spin}, so that not all choices $e^{2\pi i\,\sigma_\texttt{g}} = \pm 1$ may be allowed.

\begin{mdframed}[style=function]
    The basic object of \texttt{CasimirRFM} is a group element defined by the triple 
    \begin{equation}
        \texttt{g = \{$\D$,\{$b_1,\ldots,b_k$\},$\sigma_\gamma$\}}\,,
        \label{eq:element-def}
    \end{equation} 
    with $\D\in SL(k,\Z)$ conjugate to an orthogonal matrix, $\bvec$ defined modulo $\Z^k$, and $\sigma_\gamma\in\{0,\tfrac{1}{2}\}$ specifying the spin lift $e^{2\pi i\,\sigma_\gamma} = \pm 1$ of $\gamma\in\Gamma\subset SO(k)\to Spin(k)$.
\end{mdframed}
In the mathematical literature, and in software such as GAP \cite{GAP} and CARAT \cite{CARAT}, it is common to represent the pair $(\D,\bvec)$ in terms of the matrix 
$$\mathbf{M} = \begin{pmatrix}
    \D & \bvec \\
    0 & 1 
\end{pmatrix} \,.$$
We have included function \function{DefGenerator} that builds an object \eqref{eq:element-def} for such a matrix. 
\hypertarget{DefGenerator}{
\begin{mdframed}[style=function]
   \function{DefGenerator[M,$\sigma_\gamma$]} returns the generator \texttt{g = \{$\D$,\{$b_1,\ldots,b_k$\},$\sigma_\gamma$\}} built from the matrix \texttt{M = $\begin{pmatrix}
       \D & \bvec \\
    0 & 1 
   \end{pmatrix}$} and the choice of spin lift $\sigma_\gamma\in\{0,\frac12\}$ (set to 0 by default).
\end{mdframed}}

\subsection{Group functions}
\label{sec:group-functions}

In this subsection we describe the functions of \texttt{CasimirRFM} built for the study of finite groups $\Gamma$ that can define RFMs through the quotient \eqref{rfmdef}. For more details on the mathematical background see \cite{ValeixoBento:2025yhz}. 

To begin with, an element $\gamma$ acts on a vector $\vec{z}\in T^k$ as in \eqref{eq:affine-tranform}.

\hypertarget{ElementAction}{
\begin{mdframed}[style=function]
   \function{ElementAction[g][z]} returns the vector $\D\,\vec{z}+\bvec$ obtained by acting with element $\gamma$ on $\vec{z}$. The inputs are an element \texttt{g} of the form \eqref{eq:element-def} and a $k$--dimensional symbolic or numeric \texttt{Array}; the output is a symbolic or numeric array of the same dimension.
\end{mdframed}}

\noindent Demanding that the $\Gamma$ action has no fixed points leads to constraints on the pairs $(\mathbf{D}_\gamma,\vec{b}_\gamma)$. For instance, the equation 
\begin{equation} 
    (\mathbf{D}_\gamma-\mathbf{I})\, \vec{z}+\vec{b}_\gamma=0\,\, \text{mod}\,\,\Z^k 
    \label{fpfcond}
\end{equation}
must have no solution, so that $\vec{b}_\gamma$ must have a non-zero component along the invariant subspace of $\mathbf{D}_\gamma$, which must therefore have a non-trivial invariant subspace. 
\hypertarget{ActsFreelyQ}{
\begin{mdframed}[style=function]
   \function{ActsFreelyQ[$\Gamma$]} returns \texttt{True} if $\Gamma$ acts freely on the torus, by checking whether \eqref{fpfcond} has no solutions for non-identity elements. The input is a \texttt{List} of elements of the form \eqref{eq:element-def} and the output is a \texttt{Boolean} value.
\end{mdframed}}
Moreover, we are free to rotate $\bvec$ by any power of $\D$ and shift it by the projection of any vector $\vec{u}\in\Z^k$ onto the subspace orthogonal to the invariant subspace of $\D$, i.e. $\bvec\to\D^q + (\mathbf{I}-\D)\vec{u}$ with $q\in\Z$. The set of equivalence classes of $\bvec$ vectors is the set of solutions to the equation
\begin{equation}
    \D\,\bvec = \bvec \quad\text{mod}\quad \Lambda_{\perp} \,,
    \label{eq:allowed_shifts}
\end{equation}
where $\Lambda_{\perp}$ is the lattice defined as the projection of $\Z^k$ onto the subspace orthogonal to the invariant subspace of $\D$ \cite{ValeixoBento:2025yhz}.
\hypertarget{GetAllowedShifts}{
\begin{mdframed}[style=function]
    \function{GetAllowedShifts[$\D$]} returns the set of shift vectors $\bvec$ satisfying condition \eqref{eq:allowed_shifts} for which the resulting space-group action is fixed-point free for a generator $\D$ of the point group $\Gamma$. The input is a $(k\times k)$--dimensional numeric \texttt{Array} (typically corresponding to \texttt{g[[1]]} from \eqref{eq:element-def}) and the output is a \texttt{List} of vectors $\bvec$.
\end{mdframed}}

\noindent
Since $\Gamma$ is a \emph{finite} subgroup of isometries of $T^k$, all elements $\gamma\in\Gamma$ must have finite order, defined as the smallest integer $k$ such that $\gamma^k = \mathbf{1}$, where $\mathbf{1}$ is the identity element of $\Gamma$. In terms of the representative pair $(\D,\bvec)$, the identity is $\mathbf{1} \equiv (\mathbf{I}_k,\vec{0})$, where $\mathbf{I}_k$ is the $k\times k$ identity matrix and $\vec{0}$ the $k$-dimensional vector with all entries $0$ (mod $\Z$). 
\hypertarget{ElementOrder}{
\begin{mdframed}[style=function]
    \function{ElementOrder[g]} returns the order of \texttt{g}, i.e. the smallest integer $k$ such that $\texttt{g}^k=\mathbf{1}$. It takes \texttt{g} as an affine torus isometry, not taking into account the spin lift.
\end{mdframed}}
The finite group $\Gamma$ is generated by a set $\{\mathbf{g}_1,\mathbf{g}_2,\ldots\}$, each of which of finite order $k_i$, as 
\begin{equation}
    \Gamma = \{\mathbf{1},\mathbf{g}_1,\mathbf{g}_1^2,\ldots,\mathbf{g}_1^{k_1-1},\mathbf{g}_2,\mathbf{g}_2^2,\ldots,\mathbf{g}_2^{k_2-1},\mathbf{g}_1\mathbf{g}_2,\ldots\} \,,
    \label{eq:group}
\end{equation}
i.e. $\Gamma$ is built recursively by acting with all generators on $\Gamma_i$ and removing duplicate elements (e.g. $\mathbf{1}=\mathbf{g}_1^{k_1}=\mathbf{g}_2^{k_2}$), starting from $\Gamma_0 = \{\mathbf{1}\}$ and stopping when acting with any generator does not produce new elements.
\hypertarget{GetGroup}{
\begin{mdframed}[style=function]
    \function{GetGroup[$\{\texttt{g}_1,\texttt{g}_2,\ldots\}$]} returns the finite group generated by $\{\texttt{g}_1,\texttt{g}_2,\ldots\}$. All elements in $\{\texttt{g}_1,\texttt{g}_2,\ldots\}$ must have finite order. The input is a \texttt{List} of elements of the form \eqref{eq:element-def} and the output is another \texttt{List} of elements of the form \eqref{eq:element-def} corresponding to \eqref{eq:group}.
\end{mdframed}}

It is often convenient to exploit symmetry transformations that leave the group $\Gamma$ invariant---transformations $\vec{z}\to f(\vec{z}) = Q\,\vec{z}+\vec{q}$ that belong to the normalizer of $\Gamma$---since they become symmetries of the Casimir potential. A transformation $f(\vec{z})$, represented by the pair $(Q,\vec{q})$, lies in the normalizer of $\Gamma$ if by applying it to all elements $\gamma\in\Gamma$ we get the same group $\Gamma$. These are affine transformations of the parent $T^k$ that descend to well-defined diffeomorphisms of $\F{k}$ for which $\vec{z}$ and $f(\vec{z})$ are in the same $\Gamma$-orbit generated by \eqref{eq:affine-tranform}. Such transformations must satisfy
\begin{equation}
    f(\iota_\gamma(\vec{z})) = \iota_{\gamma'}(f(\vec{z})) \,,
\end{equation}
for some $\gamma'\in\Gamma$ and for any $\vec{z}$, leading to the pair of conditions \cite{ValeixoBento:2025yhz}
\begin{equation}
    Q\cdot\D\cdot Q^{-1} = \D[\gamma']  
    \quad\text{and}\quad 
    Q\,\bvec + (\mathbf{I} - Q\cdot\D\cdot Q^{-1})\,\vec{q} = \bvec[\gamma'] \,.
\end{equation}
A given transformation $f(\vec{z})$ may satisfy the first condition on the rotation matrices $\D$, but not the second on the shift vectors $\bvec$; when this happens, one says that the transformation normalizes the \emph{point group} that only sees rotations, but not the \emph{space group} that takes into account translations as well. In some cases, such transformations are still useful, even if they do not descend to a well-defined diffeomorphism of $\F{k}$.
\hypertarget{NormalizerQ}{
\begin{mdframed}[style=function]
    \function{NormalizerQ[$\Gamma$,\{Q,q\}]} returns \texttt{True} if the affine transformation $f(\vec{z})=Q\,\vec{z}+\vec{q}$ represented by the pair \texttt{\{Q,q\}} is in the normalizer of group $\Gamma$, i.e. if the map 
    \begin{align*}
        \D&\to Q\cdot\D\cdot Q^{-1} \\
        \bvec&\to Q\,\bvec + (\mathbf{I} - Q\cdot\D\cdot Q^{-1})\,\vec{q} 
    \end{align*}
    maps $\Gamma$ to itself. It returns \texttt{`True for point group only'} if $f(\vec{z})$ normalizes the \emph{point group}, but not the \emph{space group}, and \texttt{False} otherwise.
\end{mdframed}}

A particularly simple class of RFMs consists of those whose holonomy group $\Gamma$ is cyclic, which we refer to as \emph{cyclic} RFMs \cite{ValeixoBento:2025yhz}. Not only are these manifolds easier to analyse because the holonomy group is generated by a single element $\mathbf{g}$, they can also be classified in terms of finite-order matrices $\mathbf{A}\in GL(k,\mathbb{Z})$ representing the action of $\Gamma$ on the lattice. More precisely, one seeks matrices satisfying
\begin{align}
    \mathbf{A}^n=\mathbf{I}\,,
    \qquad
    \mathbf{A}^p\neq\mathbf{I}
    \quad\text{for all } p<n \,,
\end{align}
so that $\mathbf{A}$ has order $n$. Since two matrices related by an integral change of lattice basis define equivalent holonomy representations, the problem reduces to classifying finite-order elements of $GL(k,\mathbb{Z})$ up to integral conjugacy \cite{Ono:1988kh,ValeixoBento:2025yhz}.

The classification relies on the factorisation of $x^n-1$ into cyclotomic polynomials \cite{newman1972integral}. The $n^{\rm th}$ cyclotomic polynomial is defined as
\begin{align}
    \Phi_n(x)=
    \prod_{\substack{1\leq p\leq n\\ \gcd(p,n)=1}}
    \left(x-e^{2\pi i \frac{p}{n}}\right) \,,
\end{align}
and its roots are precisely the primitive $n$-th roots of unity---it is irreducible over $\mathbb{Q}$, has integer coefficients, and degree $\varphi(n)$ given by Euler's totient function $\varphi$. By the rational canonical form, every finite-order matrix in $GL(k,\mathbb{Z})$ is conjugate over $GL(k,\mathbb{Q})$ to a block-diagonal matrix whose blocks are companion matrices of cyclotomic polynomials \cite{CurtisReiner}, 
\begin{align}
    \mathbf{A}\sim \begin{pmatrix}
        \mathrm{C}(\Phi_{d_1}) & 0 & 0 \\
        0 & \ddots &  0\\
        0 & 0 & \mathrm{C}(\Phi_{d_q}) 
    \end{pmatrix}\,,
    \quad\text{such that}\quad
    \sum_{i} \varphi(d_i) = k \,,
\end{align}
where the companion matrix of a polynomial 
$$p(x) = x^d + a_{d-1}\,x^{d-1} + \ldots + a_1\,x + a_0$$
is defined as 
\begin{equation}
    \mathrm{C}(p) \equiv \begin{pmatrix}
        ~0~ & ~0~ & \cdots & ~0~ & -a_0 \\
        ~1~ & ~0~ & \cdots & ~0~ & -a_1 \\
        ~0~ & ~1~ & \cdots & ~0~ & -a_2 \\
        \vdots & & \ddots & & \vdots \\
        ~0~ & ~0~ & \cdots & ~1~ & a_{d-1}
    \end{pmatrix} \,.
\end{equation}
\hypertarget{GetCompanionMatrix}{
\begin{mdframed}[style=function]
    \function{GetCompanionMatrix[p(x),x]} returns the companion matrix of the polynomial \texttt{p(x)} in the form of a \texttt{SparseArray}.
\end{mdframed}}
Therefore, the rational conjugacy classes are in one-to-one correspondence with combinations of cyclotomic polynomials $\{\Phi_{d_1}\,,\,\ldots\,,\Phi_{d_q}\}$ whose degrees sum to $k$. Passing from rational to integral conjugacy is more subtle, since a given rational conjugacy class may split into several integral conjugacy classes  \cite{LatimerMacDuffee,McKee2021}. In the dimensions relevant for cyclic RFMs, these integral conjugacy classes can nevertheless be enumerated explicitly \cite{Ono:1988kh}. We can then use these block-diagonal finite-order matrices as representatives of the distinct actions of cyclic groups $\Gamma$ on the lattice. 

\hypertarget{FiniteOrderMatrices}{
\begin{mdframed}[style=function]
    \function{FiniteOrderMatrices[k]} returns block-diagonal representatives of the rational conjugacy classes of finite-order $k\times k$ integer matrices, constructed as direct sums of companion matrices of cyclotomic polynomials. The output is a list of pairs \texttt{\{order,matrix\}}.
\end{mdframed}}
Since Euler's totient function is bounded as $\phi(n)\geq \sqrt{\frac{n}{2}}$ \cite{HardyGodfreyHarold2008Aitt}, any cyclotomic polynomial $\Phi_n(x)$ contributing to a $k\times k$ matrix satisfies $n\leq 2k^2$, so that only finitely many cyclotomic companion matrices need to be considered.

\vskip 2em
\subsection{Compactification functions}
\label{sec:compactification-functions}

Recall that, by choosing the covering torus lattice $\Lambda=\Z^k$, the metric on $T^k$ and hence the one induced on the RFM is not the standard metric.
The most general metrics on the RFM are those metrics $\mathbf{G}$ on $T^k$ that satisfy \cite{ValeixoBento:2025yhz}
\begin{equation}\label{eq:moduli.fixing}
    \D^T\cdot\mathbf{G}\cdot\D = \mathbf{G} \,, \quad \forall\,\gamma\in\Gamma \,. 
\end{equation} 
Since all elements $\gamma\in\Gamma$ are generated by a given set of generators $\{\mathbf{g}_1,\mathbf{g}_2,\ldots\}$, in such a way that the matrices $\D$ are all products of $\{\D[g_1],\D[g_2],\ldots\}$, it is sufficient to check condition \eqref{eq:moduli.fixing} for the smaller set of generators. 
While the most general metric $\mathbf{G}$ on $T^k$ would contain $$\frac{k(k+1)}{2}$$ independent components, the conditions \eqref{eq:moduli.fixing} imposed by $\{\mathbf{g}_1,\mathbf{g}_2,\ldots\}$ usually reduce the number of independent components and therefore the number of moduli of the metric $\mathbf{G}$ on the RFM. From the compactification point-of-view, the group $\Gamma$ \emph{projects out} or \emph{freezes} some of the metric moduli.
\hypertarget{GetInvariantMetric}{
\begin{mdframed}[style=function]
    \function{GetInvariantMetric[$\{\texttt{g}_1,\texttt{g}_2,\ldots\}$]} returns the invariant metric $G$ and geometric moduli $\{c_1,c_2,\ldots\}$ from a set of generators $\{\texttt{g}_1,\texttt{g}_2,\ldots\}$ in the form $$\texttt{\{number of moduli,$\{c_1,c_2,\ldots\},G$\}}$$ where $G$ is a $k\times k$ symmetric matrix that is a function of the moduli $\{c_1,c_2,\ldots\}$ and satisfies \eqref{eq:moduli.fixing} for all $\gamma\in\{\texttt{g}_1,\texttt{g}_2,\ldots\}$. The input is a \texttt{List} of elements in the form \eqref{eq:element-def} and the output metric \texttt{G} is symbolic.
\end{mdframed}}

The function \hyperlink{GetInvariantMetric}{\function{GetInvariantMetric}} returns a symbolic invariant symmetric matrix, rather than an actual metric; for this matrix to be a Riemannian metric, it must also be positive-definite, which depends on the values of the parameters (moduli) $\{c_1,c_2,\ldots\}$. Therefore, not all value assignments for the moduli lead to a physical metric $G$.   

Given a $k$-dimensional RFM, we want to consider the dimensional reduction of a $D$-dimensional supersymmetric theory on the RFM, down to $d = D-k$ dimensions.
Since RFMs are quotients of tori, the kinetic terms of the moduli $\{c_1,c_2,\ldots\}$ come from the Einstein-Hilbert action on $T^k$ \cite{Etheredge:2022opl},
\begin{equation}
    \frac{1}{2\kappa_d^2}\int d^dx\, \sqrt{-g_d}\left[\frac{1}{(d-2)} (\partial\log\sqrt{\text{det}\,\mathbf{G}})^2+ \frac{1}{4}\text{Tr}((\mathbf{G}^{-1}\partial\mathbf{G})^2)\right] \,.
    \label{eq:kinetic-terms}
\end{equation}
From these we can read off the moduli space metric, which is required for instance to determine the masses of the moduli. 
\hypertarget{GetModuliModuliSpaceMetric}{
\begin{mdframed}[style=function]
    \function{GetModuliSpaceMetric[D,G,\{c$_1$,c$_2$,\ldots\},print]} returns the metric on the moduli space parameterised by the moduli \{c$_1$,c$_2$,\ldots\} of the RFM with metric \texttt{G(c$_1$,c$_2$,\ldots)}, when compactifying a \texttt{D}-dimensional theory. If \texttt{print = True}, the kinetic terms \eqref{eq:kinetic-terms} are printed (\texttt{print = False} by default). 
\end{mdframed}}
It is often convenient to reparametrise the moduli of the invariant metric (e.g. in such a way that all moduli values correspond to a physical metric $G$),
\begin{mmaCell}[
  moredefined={GetInvariantMetric,g1,g2,c1,c2,f},
  excessargument=n]{Code}
  G = GetInvariantMetric[{g1,g2,...}][[3]]/.{c1 -> f[newc],...} 
\end{mmaCell}
Therefore, the function \hyperlink{GetModuliSpaceMetric}{\function{GetModuliSpaceMetric}} takes the list of moduli as input in order to be independent of \hyperlink{GetInvariantMetric}{\function{GetInvariantMetric}} and it can be used given any parametrisation with the appropriate list of moduli \{c$_1$,c$_2$,\ldots\}.

The action of $\Gamma$ on $T^k$ also determines the cohomology classes in $H^p(\mathcal{F}_k,\R)$ from the standard basis of $H^p(T^k,\R)$, since they correspond to $p$-forms on $T^k$ that are invariant under the action \eqref{eq:affine-tranform} for all $\gamma\in\Gamma$,
\begin{equation}
    \omega_{a\ldots b}\,dz^a\wedge\ldots\wedge dz^b \to \omega_{a\ldots b}\,(\D)^{a}_{\phantom{a}a'}\ldots(\D)^b_{\phantom{b}b'}\,dz^{a'}\wedge\ldots\wedge dz^{b'} \,.
    \label{eq:form-tranform}
\end{equation}
As before, it is sufficient to check invariance under \eqref{eq:form-tranform} for the set of generators $\{\mathbf{g}_1,\mathbf{g}_2,\ldots\}$. The dimension of $H^p(\mathcal{F}_k,\R)$ is given by the Betti number $b_p$, as usual.
\hypertarget{GetInvariantForms}{
\begin{mdframed}[style=function]
    \function{GetInvariantForms[$\{\texttt{g}_1,\texttt{g}_2,\ldots\}$]} returns a basis of $H^p(\mathcal{F}_k,\R)$, i.e. a basis of $p$-forms invariant under the action \eqref{eq:form-tranform} of the group generators $\{\texttt{g}_1,\texttt{g}_2,\ldots\}$ in the form 
    \begin{equation*}
        \texttt{\{\{$1,b_1,\ldots,b_k$}\}\,,\begin{tabular}{c}
         \texttt{\{basis of $H^0(\mathcal{F}_k,\R)$,} \\
         \vdots \\
         \texttt{basis of }$H^k(\mathcal{F}_k,\R)$\texttt{\}}
    \end{tabular}\,,
    \,\texttt{basis form coefficients}\texttt{\}}
    \end{equation*}
\end{mdframed}}
The function \hyperlink{GetInvariantForms}{\function{GetInvariantForms}} outputs the basis of invariant $p$-forms in terms of linear combinations of $dz^a\wedge\ldots\wedge dz^b$ in the language of differential forms---convenient for reading off the form basis---and in component form $\omega_{a\ldots b}$ \eqref{eq:form-tranform} as a set of rules that takes into account the index antisymmetry and with which a \texttt{SparseArray} can be defined. 
The basis of $H^p(\mathcal{F}_k,\R)$ does not, in general, give a basis of the integer cohomology $H^p(\mathcal{F}_k,\Z)$ of forms with quantised periods. For flux compactifications, one must take into account flux quantisation and a properly quantised basis becomes important. 
Returning the set of rules rather than the \texttt{SparseArray} itself allows us to subsequently rescale certain basis forms to correctly implement period quantisation. 

The Casimir energies depend heavily on the spin structure chosen for the RFM \cite{ValeixoBento:2025yhz}. On the covering torus, a spin structure can be specified by the periodicity of spinor fields $\psi(\vec{z})$ under lattice transformations,
\begin{equation} 
    \psi(\vec{z}+\vec{n})=e^{2\pi i\,\vec{h}\cdot\vec{n}}\, \psi(\vec{z}) \,,
    \label{eq:spin-structure}
\end{equation}
and is encoded in a vector $\vec{h}$ whose entries are $0$ along periodic directions and $\frac12$ along anti-periodic directions. A spin structure on the torus descends to a well-defined spin structure on the RFM if it is preserved by the action of the group $\Gamma$ defining the quotient. The diffeomorphism \eqref{eq:affine-tranform} acts on spinor fields as
\begin{equation}
    \psi(\vec{z})\,\rightarrow (-1)^{2\sigma_\gamma}\mathcal{D}_{\gamma}\psi(\D\,\vec{z} + \bvec) \,,
\end{equation}
where the sign ambiguity in the spin representation is associated with the choice of spin lift $SO(k)\to\Spin(k)$, i.e. $\D\to\pm\mathcal{D}_{\gamma}\equiv e^{2\pi i\,\sigma_\gamma}\mathcal{D}_\gamma$ with $\sigma_\gamma\in \{0,\frac12\}$. A consistent spin lift of $\Gamma$ must preserve the group structure; as usual, since all group elements are obtained from applying different combinations of the generators $\{\mathbf{g}_1,\mathbf{g}_2,\ldots\}$, the choice of lifts $\{\sigma_1,\sigma_2,\ldots\}$ determines the signs $\sigma_\gamma$ of all other elements \cite{lutowski2015spin}, and a consistent choice of lift must be such that all group relations are respected.

For the spin structure on $T^k$ to be preserved by the action of $\gamma$, the vector $\vec{h}$ must satisfy \cite{ValeixoBento:2025yhz}
\begin{equation} 
    (\mathbf{I} - \D^T)\cdot \vec{h}\in\mathbb{Z}^{k} \,.
    \label{eq:spin-structure-condition-1}
\end{equation}
Moreover, for any $\D$ of order $p$, the action $\vec{z}\to\D^p\,\vec{z} + p\,\bvec$ is a translation along the vector $p\,\bvec$, which must be in the lattice $\mathbb{Z}^k$. Since the spin lift of $\D$ satisfies $(\pm\mathcal{D}_{\gamma})^p=e^{2\pi i\,(p\,\sigma_{\gamma})} \mathbf{I}$, the choice of spin structure on the covering $T^k$ along the direction $p\,\bvec$ is correlated with the choice of spin lift, 
\begin{equation} 
    p\,\sigma_\gamma\equiv \vec{h}\cdot (p\,\bvec)\quad \text{mod}\quad\mathbb{Z} \,. 
    \label{eq:spin-structure-condition-2}
\end{equation}
Note that for even $p$, we have $p\,\sigma_\gamma = 0\,\text{mod}\,\Z$ regardless of the choice of spin lift, constraining the vector $\vec{h}$ directly in such a way that $\vec{h}\cdot (p\,\bvec) = 0 \,\text{mod}\,\,\mathbb{Z}$. In contrast, when $p$ is odd the choice of lift $\sigma_\gamma$ is correlated with the choice of spin structure $\vec{h}$ for consistency.
\hypertarget{SpinStructures}{
\begin{mdframed}[style=function]
    \function{SpinStructures[\{g$_1$,g$_2$,\ldots\}]} returns an \texttt{Association} $\langle|\texttt{"GeneralForm"}\to\texttt{h}\,,$ $\texttt{"Allowed"}\to \texttt{list}|\rangle$ with the spin structures on $T^k$ that descend to a well-defined spin structure on the RFM defined as the quotient $T^k/\Gamma$ with $\Gamma$ generated by \texttt{\{g$_1$,g$_2$,\ldots\}}, according to the consistency conditions \eqref{eq:spin-structure-condition-1} and \eqref{eq:spin-structure-condition-2}. The \texttt{"GeneralForm"} key points to the most general consistent structure \texttt{\{h$_1$,\ldots,h$_k$\}}; the \texttt{"Allowed"} key points to a list of all allowed spin structures (with $h_i = 0$ for periodic boundary conditions and $h_i=\frac12$ for anti-periodic boundary conditions around direction $i$ on the torus).
\end{mdframed}}

\subsection{Casimir energies}
\label{sec:Casimir}

Upon compactification of a $D$-dimensional field theory on a $k$-dimensional RFM, the lower-dimensional Casimir energy, i.e. the effective potential coming from the Casimir term, is given by \cite{ValeixoBento:2025yhz}
\begin{equation} 
    \Vcas 
    = -\frac{\Gamma(s)}{2\pi^{s}} \frac{1}{\vert\Gamma\vert}\sum^{\sim}_{\substack{\gamma\in\Gamma\\\vec{n}\in\mathbb{Z}^k}}  \sum_{\textbf{r}}\int_{[0,1]^k} d^k\vec{z}\,\sqrt{G}\,  \frac{\Tr{\bf r}{\D}\, e^{2\pi i\,\vec{h}\cdot\vec{n}} }{\vert\vec{z}- (\mathbf{D}_\gamma\,  \vec{z} + \vec{b}_\gamma+\vec{n})\vert^D} \,.
    \label{vcasfed}
\end{equation}
The integral over the compact space $\mathcal{F}_k$ is expressed as an integral over the covering torus $T^k$, with fundamental domain $[0,1)^k$ over which we integrate and equipped with a metric $\mathbf{G}$ satisfying \eqref{eq:moduli.fixing}; the integral is divided by the order $\vert \Gamma\vert$ of the finite group $\Gamma$ defining the RFM because the torus is covering $\mathcal{F}_k$ a total of $\vert \Gamma\vert$ times. The $\sim$ over the sum indicates that we sum over all $\gamma\in\Gamma$ and $\vec{n}\in\Z^k$ such that the denominator does not vanish, which only happens for $\gamma=\mathbf{I}\,$, $\vec{n}=0$ due to the fixed-point-freeness of the group. We also sum over all representations $\mathbf{r}$ present in the massless spectrum of the $D$-dimensional theory. 
The vector $\vec{h}$ encodes twisted boundary conditions of various fields, that transform as a phase under translations. The $\Z_2$-twisted boundary conditions of fermion fields, related to the spin structures discussed in the previous section, are a particular example; however, the function \function{Ewald} accepts more general phase vectors, allowing for other twisted boundary conditions.

The potential \eqref{vcasfed} can be expressed as a sum---over the elements $\gamma\in\Gamma$ and the massless representations $\mathbf{r}$---of infinite lattice sums, such that each $\gamma\in\Gamma$ gives a definite contribution $\mathcal{E}(\gamma)$ \cite{ValeixoBento:2025yhz},
\begin{equation}
    \Vcas = \sum_{\gamma\in\Gamma}\sum_{\textbf{r}} \Tr{\bf r}{\D}\,\mathcal{E}(\gamma)
    \,,\quad
    \mathcal{E}(\gamma) = -\hat{\delta}_{\vec{h}}\,\frac{\Gamma(s_\gamma)}{2\pi^{s_\gamma}}\cdot
    \frac{\sqrt{G_\parallel}}{\vert\Gamma\vert} \sum_{\vec{\xi}\in\,\Xi_\gamma} \frac{e^{2\pi i \,\Vec{\beta}_\gamma\cdot\vec{\xi}}}{|\vec{\xi} + \bvec^{\,\parallel}|^{2s_\gamma}_\parallel} \,.
    \label{eq:Casimir-general}
\end{equation}
In this formula, $s_\gamma = s - \frac{k-k'}{2}$, where $s=D/2$ and $k'$ is the dimension of the subspace invariant under $\D$. The lattice $\Xi_\gamma$ and the vector $\vec{b}_\gamma^{\,\parallel}$ are the projections of the lattice $\Z^k$ and vector $\bvec$ onto the invariant subspace of $\D$, while $\mathbf{G}_\parallel$ is the metric induced on this subspace and used for the inner product $|\cdot|_\parallel$. 
Each $\mathcal{E}(\gamma)$ behaves exactly as the contribution to the potential that would arise from an effective $k'$--dimensional ``Casimir brane'' wrapped on the invariant subspace of $\D$ and with a tension given by \eq{eq:Casimir-general} \cite{ValeixoBento:2025yhz}. 
\hypertarget{CasimirBranes}{
\begin{mdframed}[style=function]
    \function{CasimirBranes[$\texttt{g}$]} returns the position of the Casimir branes associated with \texttt{g} corresponding to the locus in $T^k$ of the invariant subspace under $\D$, i.e. the subspace defined by the solutions to $\D\,\vec{z}=\vec{z}\in T^k$. 
\end{mdframed}}
Finally, the factor $\hat{\delta}_{\vec{h}}$ evaluates to $1$ if there exists a solution to the equation 
\begin{equation}
    (\mathbf{I} - \D)^T\vec{\eta} = (\mathbf{I} - \D)^T\vec{h} \,,
    \label{eq:eta-condition}
\end{equation}
with $\vec{\eta}\in\Z^k$, and zero otherwise, in which case the sum $\mathcal{E}(\gamma)$ vanishes identically. Interestingly, the factor $\hat{\delta}_{\vec{h}}$ is related to a field theory version \cite{ValeixoBento:2025yhz} of Atkin-Lehner symmetry \cite{atkin1970hecke,Moore:1987ue,Dienes:1990qh}. Note that $\hat{\delta}_{\vec{h}}=1$ regardless of $\D$ whenever $\vec{h}=\vec{0}$, e.g. for bosons with fully periodic boundary conditions on the covering torus. Whenever there is a solution $\vec{\eta}\in\Z^k$, the vector $\vec{\beta}_\gamma$ in \eqref{eq:Casimir-general} is defined as $\vec{\beta}_\gamma \equiv \vec{h} - \vec{\eta}$ and belongs to the invariant subspace of $\D^T$.

The lattice sums in \eqref{eq:Casimir-general} are computed using the Ewald summation formula (see Appendix C of \cite{ValeixoBento:2025yhz} for derivation and details),
\begin{align}
    \sum_{\vec{n}+\vec{c}\neq0}\frac{e^{2\pi i \,\vec{h}\cdot\vec{n}}}{\vert \vec{n}+\vec{c}\vert^{2s}} 
    =& \sum_{\vec{n}+\vec{c}\neq0}\frac{e^{2\pi i \vec{h}\cdot\vec{n}}}{\vert \vec{n}+\vec{c}\vert^{2s}} \frac{\Gamma(s,\alpha \vert \vec{n}+\vec{c}\vert^{2})}{\Gamma(s)} \nonumber \\
    &+ \frac{\pi^{2s-\frac{k}{2}}}{\sqrt{G}} \sum_{\vec{k}-\vec{h}\neq0}\,\frac{e^{2\pi i(\vec{k}-\vec{h})\cdot\vec{c}}}{ \vert\vec{k}-\vec{h}\vert_\mathrm{D}^{k-2s}}\,\frac{\Gamma\left(\tfrac{k}{2}-s,\tfrac{\pi^2}{\alpha}\,\vert\vec{k}-\vec{h}\vert_\mathrm{D}^2\right)}{\Gamma(s)} \nonumber \\
    &+ \delta_{\vec{h},\vec{0}}\, \frac{\pi^{\frac{k}{2}}\alpha^{s-k/2}}{\sqrt{G}\, \Gamma(s)\left(s-\frac{k}{2}\right)}-\frac{\alpha^s\, e^{-2\pi i \vec{h}\cdot\vec{c}}}{\Gamma(s+1)}\,\chi_{\mathbb{Z}^k}(\vec{c}) \,.
    \label{ewald} 
\end{align}
where $\vert \vec{v}\vert_\mathrm{D}^2\equiv G_{ij}^{-1} v^i v^j$ is the canonical norm in momentum space. 
This expression is exact for any $\alpha$, but in practice one truncates both the sum over $\vec{n}$ (position space) and the one over $\vec{k}$ (momentum space), and chooses $\alpha$ to maximize the speed of convergence. 

\hypertarget{Ewald}{
\begin{mdframed}[style=function]
    \function{Ewald[c,h,s,$\alpha$,G,$\varepsilon$]} numerically computes the lattice sum \eqref{ewald}, with vector norms computed using the metric \texttt{G}, shift vector $\vec{c}$ and phase vector $\texttt{h}=\{h_1,\ldots,h_k\}$; the parameter $\alpha$ specifies the split between real and reciprocal sums; 
    the parameter $\varepsilon$ controls the estimated truncation of the numerical sums such that $\varepsilon = 10^{-p}$ targets roughly $p$ decimal digits ($\varepsilon=10^{-4}$ by default). 
    All input arguments are numeric and the output is the numeric result of the sum.
\end{mdframed}}

In our numerical implementation, we first determine the set of vectors with norm smaller than a given cutoff, i.e. within the ellipsoid defined by the metric and the cutoff, and perform the sum only after this set of vectors has been fully identified and stored in memory. This provides a large speedup over implementations that sum over a large square box of points, which despite not needing to spend computer resources identifying the appropriate set for the sum, spend time evaluating many points of very large norm with negligible contribution. This is particularly important in larger dimension where there are many points in the square box that are not contained in the ellipsoid for a generic norm. 
In the private function \function{GetEwaldLatticeByNorm[$r^2$,G,$\vec{\texttt{c}}$]} we implement a recursive dimensional-reduction method inspired by the Fincke–Pohst algorithm for short-vector enumeration in a lattice \cite{fincke1985improved}:
rather than scanning a rectangular box, it recursively bounds each coordinate using the positive-definite quadratic form. Our implementation uses Schur complements for the recursive reduction and does not require a separate lattice-reduction stage.

More precisely, this function generates the lattice of vectors with 
\begin{equation}
    |\vec{n}+\vec{c}|_G^2 < r^2 \,,
    \label{eq:original-ellipsoid}
\end{equation}
with inner-product given by the metric $G$. Decomposing the vectors $\vec{n}=(n_1,\vec{n}')$ and $\vec{c}=(c_1,\vec{c}')$, and the metric
\begin{equation}
    G = \begin{pmatrix}
        G_{11} & \vec{G}_1^T \\ 
        \vec{G}_1 & \mathrm{G}'
    \end{pmatrix} \,,
\end{equation}
we can rewrite \eqref{eq:original-ellipsoid} as 
\begin{equation}
    (n_1 + c_1)^2 [G_{11} - \vec{G}_1^T(\mathrm{G}')^{-1}\vec{G}_1] + |\vec{n}' + \vec{c}' + (n_1+c_1)(\mathrm{G}')^{-1}\vec{G}_1|_{\mathrm{G}'}^2 < r^2 \,. 
\end{equation}
Since the second term is also a non-negative norm, this implies
\begin{equation}
    n_1 \in \left[-\sqrt{\frac{r^2}{G_{11} - \vec{G}_1^T(\mathrm{G}')^{-1}\vec{G}_1}} - c_1 \,,\, \sqrt{\frac{r^2}{G_{11} - \vec{G}_1^T(\mathrm{G}')^{-1}\vec{G}_1}} - c_1\right]\cap\Z \,.
\end{equation}
Then, for each value of $n_1$ in this range, the remaining coordinates must satisfy the same constraint \eqref{eq:original-ellipsoid} for the redefined quantities
\begin{align*}
    \vec{n}\to\vec{n}' \,,
    \quad\vec{c}\to\vec{c}' + (n_1+c_1)(\mathrm{G}')^{-1}\vec{G}_1 \,, 
    \quad G\to\mathrm{G}' \,,
\end{align*}
one dimension lower than in the original problem. This allows us to solve the problem recursively, one direction at a time, reducing the problem by one dimension at each step, until we fill in the ellipsoid with the points of interest. 

Now the cutoff for each sum in \eqref{ewald} that determines the accuracy of the numerical summation can be chosen by bounding the contribution from all the terms with norm larger than some $r_\mathrm{max}^2$. For example, for the first sum we want to find $r_\mathrm{max}^2$ such that  
\begin{equation*}
    \sum_{|\vec{n}+\vec{c}|_G^2 > r_\mathrm{max}^2}\frac{e^{2\pi i \vec{h}\cdot\vec{n}}}{\vert \vec{n}+\vec{c}\vert^{2s}_G} \frac{\Gamma(s,\alpha \vert \vec{n}+\vec{c}\vert^{2}_G)}{\Gamma(s)} < \frac{\varepsilon}{2} \,,
\end{equation*}
for a given accuracy $\varepsilon$ (split evenly between the two sums). For large enough norms, we can approximate the sum by an integral over $\rho = r^2$,
\begin{align}
    \sum_{|\vec{n}+\vec{c}|_G^2 > r_\mathrm{max}^2}\frac{e^{2\pi i \vec{h}\cdot\vec{n}}}{\vert \vec{n}+\vec{c}\vert^{2s}_G} \frac{\Gamma(s,\alpha \vert \vec{n}+\vec{c}\vert^{2}_G)}{\Gamma(s)}
    &\approx \frac{\pi^{k/2}}{\Gamma(k/2)}\frac{1}{\sqrt{G}}\int_{r_\mathrm{max}^2}^\infty \,d\rho\,\rho^{\frac{k}{2}-s-1} \frac{\Gamma(s,\alpha\,\rho)}{\Gamma(s)}  \\
    &= \frac{\pi^{k/2}}{\Gamma(k/2)}\frac{1}{\sqrt{G}}
        \frac{\alpha^{s-\frac{k}{2}}\Gamma(\frac{k}{2},\alpha\,r_{\mathrm{max}}^2) - (r_{\mathrm{max}}^2)^{\frac{k}{2}-s} \Gamma(s,\alpha\,r_{\mathrm{max}}^2)}{\left(\frac{k}{2}-s\right)\Gamma(s)} \nonumber \,,
\end{align}
where we have included a volume factor that estimates the number of points with a given norm,
\begin{equation}
    dV = \frac{2\pi^{k/2}}{\Gamma(k/2)}\,\frac{r^{k-1}}{\sqrt{G}}\,dr = \frac{\pi^{k/2}}{\Gamma(k/2)}\,\frac{\rho^{\frac{k}{2}-1}}{\sqrt{G}}\,d\rho \,.
    \label{eq:volume-factor}
\end{equation}
The result is given in terms of powers and $\Gamma$-functions, and is monotonically decreasing as $r_{\mathrm{max}}$ increases; it can therefore be efficiently solved by Mathematica's \texttt{FindRoot} solver. This is what we implement in the package. Likewise, for the reciprocal sum 
\begin{equation}
    \frac{\pi^{2s-\frac{k}{2}}}{\Gamma(s)\sqrt{G}} \sum_{|\vec{h}-\vec{k}|^2_{D} > \Tilde{r}_{\mathrm{max}}^2}\, \Gamma \left(\frac{k}{2}-s,\frac{\pi ^2 \vert\vec{h}-\vec{k}\vert_D^2}{\alpha }\right) \frac{e^{-2\pi i(\vec{h}-\vec{k})\cdot\vec{c}}}{ \vert\vec{h}-\vec{k}\vert_D^{k-2s}} < \frac{\varepsilon}{2} \,,    
\end{equation}
we compute the integral
\begin{align}
    &\frac{\pi^{2s}}{\Gamma(k/2)}\int_{\Tilde{r}_{\mathrm{max}}^2}^\infty d\rho\,\rho^{s-1}\,\frac{\Gamma(\tfrac{k}{2} - s,\tfrac{\pi^2\,\Tilde{r}_{\mathrm{max}}^2}{\alpha})}{\Gamma(s)} 
    = \frac{\alpha^2\Gamma(\tfrac{k}{2},\tfrac{\pi^2\Tilde{r}_{\mathrm{max}}^2}{\alpha}) - \pi^{2s}(\Tilde{r}_{\mathrm{max}}^2)^s\,\Gamma(\tfrac{k}{2}-s,\tfrac{\pi^2\Tilde{r}_{\mathrm{max}}^2}{\alpha})}{\Gamma(k/2)\,s\,\Gamma(s)} \,, 
\end{align}
and as before find $\Tilde{r}_{\mathrm{max}}^2$ using Mathematica's \texttt{FindRoot}. Note that the factor of $\sqrt{G}$ in the reciprocal sum cancels the factor of $\sqrt{G}$ that arises from the volume element \eqref{eq:volume-factor} when the inverse metric $G^{-1}$ is used.

The  sums over $\vec{n}$ and $\vec{k}$ are performed using compiled code (\texttt{CompilationTarget$\to$"C"}, \texttt{RuntimeOptions$\to$"Speed"}) which significantly increases the speed of summation. In order to be compiled, our implementation uses \texttt{\$MachinePrecision} rather than Mathematica's software-implemented arbitrary precision arithmetic. This substantially improves performance, at the cost of restricting the numerical evaluation to approximately 16 decimal digits.

The contribution of element $\gamma\in\Gamma$ to the Casimir potential only requires the evaluation of a sum over the sublattice invariant under $\D$ \eqref{eq:Casimir-general}; for non-trivial elements, this lattice is smaller than the full $\Z^k$ lattice. Therefore, before performing the sums numerically, we project onto this invariant sublattice $\Xi_{\gamma}$ and evaluate $\delta_{\vec{h}}$. 
\hypertarget{ReducedLatticeSum}{
\begin{mdframed}[style=function]
    \function{ReducedLatticeSum[D,g,h,G,$\varepsilon$]} evaluates numerically the lattice sum $\mathcal{E}(\texttt{g})$ corresponding to element \texttt{g} \eqref{eq:Casimir-general}\footnote{For convenience, the output is $\left(\frac{\Gamma(s)}{2\pi^s}\frac{1}{|\Gamma|}\right)^{-1}\mathcal{E}(\texttt{g})$, rather than $\mathcal{E}(\texttt{g})$ itself.}, with $s=\texttt{D}/2$, vector norms computed with metric \texttt{G} and phase vector $\texttt{h}=\{h_1,\ldots,h_k\}$; 
    the parameter $\varepsilon$ controls the estimated truncation of the numerical sums such that $\varepsilon = 10^{-p}$ targets roughly $p$ decimal digits ($\varepsilon=10^{-4}$ by default);
    this function uses the element \texttt{g} and phase vector \texttt{h} to evaluate the factor $\delta_{\vec{h}}$ through \eqref{eq:eta-condition} before calling \hyperlink{Ewald}{\function{Ewald}} for summation.
\end{mdframed}}
This function is already adapted to the context of Casimir potentials, taking the dimension $\texttt{D}=2s$ as input rather than the parameter $s$. 

In order to compute the full Casimir potential $\Vcas$ \eqref{eq:Casimir-general}, we still need to include the numerical factors, including the trace $\Tr{r}{\D}$ in each representation $\mathbf{r}$ in the massless spectrum of the theory, and to sum over all representations and all elements $\gamma\in\Gamma$.
\hypertarget{TracedSum}{
\begin{mdframed}[style=function]
    \function{TracedSum[D,g,spectrum,hspin,G,\{$\phi_i\to\overline{\phi}_i$\},$\varepsilon$]} uses \hyperlink{ReducedLatticeSum}{\function{ReducedLatticeSum}} to compute the lattice sums numerically and sums over all representations in \texttt{spectrum} taking into account the traces $\Tr{r}{\D[g]}$ of element \texttt{g} \eqref{eq:Casimir-general}\footnote{For convenience, the output is $\left(\frac{\Gamma(s)}{2\pi^s}\frac{1}{|\Gamma|}\right)^{-1}\sum_{\textbf{r}}\Tr{r}{\D[g]}\,\mathcal{E}(\texttt{g})$, rather than $\sum_{\textbf{r}}\Tr{r}{\D[g]}\,\mathcal{E}(\texttt{g})$ itself.}; the sums are evaluated with $s=\texttt{D}/2$, and vector norms computed with metric \texttt{G}$(\phi_i)$ evaluated at the point in moduli space given by the replacement rule \{$\phi_i\to\overline{\phi}_i$\}; \texttt{hspin} is the vector encoding the spin structure on the RFM; the phase vector $\vec{h}$ in each sum is determined by the \texttt{"TwistedBoundaryConditions"} key for each field in \texttt{spectrum} (by default it is $\vec{h}=\vec{0}$ for bosons and \texttt{hspin} for fermions);
    the parameter $\varepsilon$ controls the estimated truncation of the numerical sums such that $\varepsilon = 10^{-p}$ targets roughly $p$ decimal digits ($\varepsilon=10^{-4}$ by default).
\end{mdframed}}
The replacement rules \{$\phi_i\to\overline{\phi}_i$\} specifying the point in moduli space must evaluate \texttt{G} to a numerical, real and positive-definite matrix.
The traces $\Tr{r}{\D[g]}$ are computed using the functions in Table \ref{tb:list-functions-traces}, which we discuss in more detail in the next section. 
The \texttt{spectrum} of the theory is specified through the \texttt{Association}
\begin{align*}
    \texttt{spectrum} &= \langle|
    \texttt{"Bosons"}\to\texttt{bosonList}\,,\,
    \texttt{"Fermions"}\to\texttt{fermionList}|
    \rangle \,,
\end{align*}
where \texttt{bosonList} and \texttt{fermionList} are lists of \texttt{Associations},
\begin{align*}
    \texttt{bosonList} &= 
    \{
        \langle|\texttt{"Type"}\to\ldots \,,
        \texttt{"Multiplicity"}\to\ldots \,, \texttt{"Rank"}\to\ldots \,, 
        \texttt{"SelfDual"}\to\ldots 
        |\rangle
    \} \\
    \texttt{fermionList} &= 
    \{
        \langle|\texttt{"Type"}\to\ldots \,,
        \texttt{"Multiplicity"}\to\ldots \,, 
        \texttt{"Weyl"}\to\ldots 
        |\rangle
    \} \,.
\end{align*}
Valid values for \texttt{"Type"} are \texttt{\{"Graviton","Form","Scalar","Spinor","Gravitino"\}}; \texttt{"Multiplicity"} specifies how many fields of this type are in the spectrum (\texttt{"Multiplicity"$\to$1} by default); for $p$-forms, \texttt{"Rank"}$\to p$ specifies the rank of the form; \texttt{"SelfDual"$\to$True} indicates a self-dual form (\texttt{"SelfDual"$\to$False} by default); for fermions \texttt{"Weyl"$\to$True} indicates that the fermion is in the Weyl (definite-chirality) representation (\texttt{"Weyl"$\to$False} by default; option ignored in odd dimensions, since Weyl spinors do not exist). All types also admit the option \texttt{"TwistedBoundaryConditions"}$\to\{h_1,\ldots,h_k\}$, encoding transformation by a phase under translations on the covering torus; the default for \texttt{"Bosons"} is $\{0,\ldots,0\}$ for fully periodic boundary conditions, while for \texttt{"Fermions"} it is fixed by the spin structure on the RFM. To access only bosons use \texttt{KeyTake[spectrum,"Bosons"]} and to access only fermions use \texttt{KeyTake[spectrum,"Fermions"]}.
For example, the spectrum of M-theory is specified as follows
\begin{align*}
    \langle|
    &\texttt{"Bosons"}\to\{
        \langle|\texttt{"Type"}\to\texttt{"Graviton"} 
        |\rangle\,, 
        \langle|\texttt{"Type"}\to\texttt{"Form"} \,, \texttt{"Rank"}\to 3 
        |\rangle
    \}\,,\, \\
    &\texttt{"Fermions"}\to\{
        \langle|\texttt{"Type"}\to\texttt{"Gravitino"}
        |\rangle
    \}|
    \rangle \,,
\end{align*}
while the spectrum of Type IIB supergravity would be specified as
\begin{align*}
    \langle|
    \texttt{"Bosons"}&\to\{
        \langle|\texttt{"Type"}\to\texttt{"Graviton"} 
        |\rangle \,, \\
        &\phantom{\to\{}~\langle|\texttt{"Type"}\to\texttt{"Scalar"} \,, \texttt{"Multiplicity"}\to 2 
        |\rangle \,, \\
        &\phantom{\to\{}~\langle|\texttt{"Type"}\to\texttt{"Form"} \,, \texttt{"Rank"}\to 2 \,, \texttt{"Multiplicity"}\to 2  
        |\rangle \,, \\
        &\phantom{\to\{}~\langle|\texttt{"Type"}\to\texttt{"Form"} \,, \texttt{"Rank"}\to 4 \,, \texttt{"SelfDual"}\to \texttt{True}
        |\rangle
    \}\,,\, \\
    \texttt{"Fermions"}&\to\{
        \langle|\texttt{"Type"}\to\texttt{"Spinor"} \,, \texttt{"Weyl"}\to\texttt{True} \,, \texttt{"Multiplicity"}\to 2  
        |\rangle \,, \\
        &\phantom{\to\{}~\langle|\texttt{"Type"}\to\texttt{"Gravitino"} \,, \texttt{"Weyl"}\to\texttt{True} \,, \texttt{"Multiplicity"}\to 2 
        |\rangle
    \}|
    \rangle \,.
\end{align*}
The \texttt{"Weyl"} key selects a definite-chirality representation, but does not distinguish between chiralities. This is enough for the computation of traces in Riemann-flat compactifications because every holonomy element of a Riemann-flat manifold has a $+1$ eigenvalue, and the traces of the two different-chirality spinor representations coincide on the holonomy group. 

\vskip 1em
\noindent 
We are finally able to introduce the two key functions of \texttt{CasimirRFM}. The full potential $\Vcas$ \eqref{eq:Casimir-general} can be obtained directly using the function \hyperlink{CasimirPotential}{\function{CasimirPotential}}.

\hypertarget{CasimirPotential}{
\begin{mdframed}[style=function]
    \function{CasimirPotential[D,$\Gamma$,spectrum,hspin,G,\{$\phi_i\to\overline{\phi}_i$\},$\varepsilon$]} computes the Casimir potential $\Vcas$ \eqref{eq:Casimir-general} summing over all elements $\gamma\in\Gamma$, using \hyperlink{TracedSum}{\function{TracedSum}} for all lattice sums, including traces over all representations in the \texttt{spectrum} and all numerical factors; \texttt{hspin} is the vector encoding the spin structure on the RFM; the phase vector $\vec{h}$ in each sum is determined by the \texttt{"TwistedBoundaryConditions"} key for each field in \texttt{spectrum} (by default it is $\vec{h}=\vec{0}$ for bosons and \texttt{hspin} for fermions);
    the parameter $\varepsilon$ controls the estimated truncation of the numerical sums such that $\varepsilon = 10^{-p}$ targets roughly $p$ decimal digits ($\varepsilon=10^{-4}$ by default).
\end{mdframed}}
It is also useful to study the Casimir energy density $\rho_{\rm Cas}(\vec{z})$ in $D$ dimensions, rather than the $(D-k)$-dimensional Casimir potential (e.g. to visualize Casimir brane profiles or to estimate the average gradients). This can be computed at a given point $\vec{z}\in T^k$ directly by evaluating the integrand of \eqref{vcasfed}, while still summing over all elements $\gamma\in\Gamma$ and all representations $\mathbf{r}$ in the massless spectrum of the theory.

\hypertarget{CasimirEnergyDensity}{
\begin{mdframed}[style=function]
    \function{CasimirEnergyDensity[D,$\Gamma$,spectrum,hspin,G,\{$\phi_i\to\overline{\phi}_i$\},z,$\varepsilon$]} computes the Casimir energy density $\rho_{\rm Cas}(\vec{z})$ summing over all elements $\gamma\in\Gamma$, using \hyperlink{Ewald}{\function{Ewald}} for all lattice sums, including traces over all representations in the \texttt{spectrum} and all numerical factors; \texttt{hspin} is the vector encoding the spin structure on the RFM; the phase vector $\vec{h}$ in each sum is determined by the \texttt{"TwistedBoundaryConditions"} key for each field in \texttt{spectrum} (by default it is $\vec{h}=\vec{0}$ for bosons and \texttt{hspin} for fermions);
    the parameter $\varepsilon$ controls the estimated truncation of the numerical sums such that $\varepsilon = 10^{-p}$ targets roughly $p$ decimal digits ($\varepsilon=10^{-4}$ by default).
\end{mdframed}}
The definition of \function{CasimirEnergyDensity} does not include the factor of $1/|\Gamma|$ that appears in the Casimir potential \eqref{vcasfed}, since this factor arises from expressing the integral over the RFM by an integral over the covering torus. 

\subsection{Lorentz group traces under different representations}
\label{sec:traces}

The Casimir energy involves traces over representations of the group action on the Hilbert space of the massless particles in $D$ dimensions. The little group acting on $D$-dimensional massless particles is $SO(D-2)$, which decomposes as
\begin{equation}
    SO(D-2)\to SO(k)\times SO(d-2) \,,
\end{equation}
as we compactify the theory on a $k$-dimensional manifold down to $d=D-k$ dimensions, and accurately counts the physical polarizations (degrees of freedom) for each field.
The geometric action we use to obtain a Riemann-flat manifold as a quotient of $T^k$ is in the $SO(k)$ factor, but it is useful to work in terms of $SO(D-2)$ matrices. Hence, we will embed the $SO(k)$ matrix in the vector representation as a block-diagonal $SO(D-2)$ matrix $\mathbf{M}$, with an additional $(d-2)\times(d-2)$ block representing the identity. 

In principle, the matrix $\mathbf{M}$ must be orthogonal. However, because the trace of a matrix is invariant under similarity transformations, $\mathbf{M}$ can be replaced by $\mathbf{X} \mathbf{M}\mathbf{X}^{-1}$ for any non-singular matrix $\mathbf{X}$. In particular, we may replace the orthogonal matrices by non-orthogonal representatives of themselves (for instance, in terms of the $SL(k,\mathbb{Z})$ matrix $\D$ that implements the action in the lattice basis).

The trace in the symmetric-traceless tensor representation (i.e. the graviton) is given by
\begin{equation}
    \Tr{}{\mathbf{M}^{\text{Sym.}}}=\frac12\left(\Tr{}{\mathbf{M}}^2 + \Tr{}{\mathbf{M}^2}\right)-1 \,,
\end{equation}
where the $-1$ removes the trace (see Appendix F of \cite{ValeixoBento:2025yhz} for details).
\hypertarget{TraceSym}{
\begin{mdframed}[style=function]
    \function{TraceSym[g,D]} returns the trace of element \texttt{g} in the symmetric-traceless (graviton) representation of $SO(D-2)$. 
\end{mdframed}}

The trace in the $p$-index antisymmetric representation (i.e. $p$-form) can be extracted from the general formula \cite{vanRitbergen:1998pn} encoding the character of a fully antisymmetric representation\footnote{Note a typo in the rhs of (49) of \cite{vanRitbergen:1998pn}, the term $\text{Ch}_R(lF)$ should be divided by $l$ \cite{weyl1946classical,Schellekens:1986xh}.} via a generating function,
\begin{equation} 
    \sum_{k=0}^\infty x^k\Tr{}{\mathbf{M}^{k\text{-Antisym.}}}= \prod_{l=1}^\infty\exp\left(-(-x)^l\frac{\text{Tr}(\mathbf{M}^l)}{l}\right) \,,
    \label{eq:character-traces}
\end{equation}
and it is given by the coefficient of $x^p$ in its Taylor series expansion.
\hypertarget{TraceForm}{
\begin{mdframed}[style=function]
    \function{TraceForm[p,g,D]} returns the trace of element \texttt{g} in the antisymmetric $p$-form representation of $SO(D-2)$ by picking the coefficient of $x^p$ in the expansion of \eqref{eq:character-traces}. 
\end{mdframed}}

The vector-spinor representation---a tensor product between a vector and a spinor---decomposes into a gamma-traceless Rarita-Schwinger (gravitino) representation and a spinor representation. Consequently, the trace in the gravitino representation can be obtained by subtracting the spinor trace from the trace in the vector-spinor representation,
\begin{equation} 
    \Tr{}{\mathbf{M}^{\text{R.S.}}} = \left(\Tr{}{\mathbf{M}} - 1 \right)\Tr{}{\mathbf{M}^\text{Spinor}} \,.
    \label{rs}
\end{equation}

\hypertarget{TraceRS}{
\begin{mdframed}[style=function]
    \function{TraceRS[g,D,Weyl]} returns the trace of element \texttt{g} in the vector-spinor (Rarita-Schwinger/gravitino) representation of $SO(D-2)$, taking into account the spin lift of \texttt{g}; the parameter \texttt{Weyl} must be set to \texttt{True} for Weyl spinors, and \texttt{False} otherwise (\texttt{Weyl = False} by default).
\end{mdframed}}

For traces in the spinor representation it is more convenient to use Cartan decomposition and encode the representations in terms of weight vectors as follows. 
The Cartan subalgebra of $\mathfrak{so}(D-2)$, defined as a maximally commuting set of generators, corresponds to a set of rotations in mutually orthogonal planes.
We can write any $SO(D-2)$ element in the form
\begin{align}
    \mathbf{M} = \exp\left(i \theta_\alpha E^3_\alpha \right) \,,
\end{align}
where $E^3_\alpha$ is a diagonal generator that is a linear combination of the Cartan generators, i.e. $E^3_\alpha = |\alpha|^{-2}(\alpha\cdot H)$, with $\alpha$ a root of $SO(D-2)$ and $H_i$ the Cartan generators. The coefficients $\theta_\alpha$ correspond to the eigenvalues of the matrix $\mathbf{M}$, i.e. the element in the fundamental/vector representation of $SO(D-2)$, encoding the angle of the rotation along each mutually orthogonal plane.
Since we can decompose each representation into eigenstates $|\mu\rangle$ of $H_i$ labelled by a weight vector $\vec{\mu}$, once we know each decomposition in terms of its weights we can take the trace in some representation $R$ as
\begin{align}
    \Tr{R}{\mathbf{M}} = \sum_{\vec{\mu}\in R} \exp\left(i\, \vec{\theta}\cdot\vec{\mu}\right) \,,
    \label{eq:trace-formula}
\end{align}
where $\vec{\mu}\in R$ includes all weight vectors $\vec{\mu}$ in the representation $R$ (e.g. there are $(D-2)$ such vectors in the fundamental representation and $\tfrac{(D-1)(D-2)}{2}-1$ in the rank-2 symmetric traceless representation). We then find that the traces computed in this way match the ones computed using the character formulas above. For the weight decomposition of each representation, it matters whether $D-2$ is even or odd, in other words whether the group is $SO(2n)$ or $SO(2n+1)$.

\subsection*{\underline{$SO(2n)~(D\text{ even})$}}
\vspace{1em}
\begin{itemize}
    \item \textbf{Fundamental (vector)} representation has weight decomposition given by all permutations of $n$-dimensional vectors $(\pm1,0,\ldots,0)$, a total of $2n$ weights and a representation of dimension $2n$. 

    \item \textbf{Symmetric traceless (graviton)} representation has weight decomposition given by all permutations of $n$-dimensional vectors $(\pm2,0,\ldots,0)$, $(1,\pm1,0,\ldots,0)$, together with $(0,\ldots,0)$ with multiplicity $n-1$, a total of $2n(2n+1)/2 - 1$ weights. 

    \item \textbf{Antisymmetric rank$-k$} representation has weight decomposition given by all sums $\mu_1 + \ldots + \mu_k$ of distinct weights $\mu_i$ in the fundamental (vector) representation, i.e. of $n$-dimensional vectors $(\pm1,0,\ldots,0)$, for a total of $\binom{2n}{k}$ weights. 

    \item \textbf{Spinor} representation has weight decomposition given by all $n$-dimensional vectors of the form $\tfrac{1}{2}(\pm1,\ldots,\pm1)$, with dimension $2^n$. This representation decomposes into two irreducible representations of definite chirality determined by the parity of minus signs; therefore, in even dimensions $D$ there exist Weyl spinors of definite chirality of dimension $2^{n-1}$. 
\end{itemize}

\subsection*{\underline{$SO(2n+1)~(D\text{ odd})$}}

\begin{itemize}
\vspace{1em}
    \item\textbf{Fundamental (vector)} representation has weight decomposition given by all permutations of $n$-dimensional vectors $(\pm1,0,\ldots,0)$ plus the vector $(0,\ldots,0)$, a total of $2n+1$ weights and a representation of dimension $2n+1$. 

    \item\textbf{Symmetric traceless (graviton)} representation has weight decomposition given by all permutations of $n$-dimensional vectors $(\pm2,0,\ldots,0)$, $(1,\pm1,0,\ldots,0)$, $(\pm1,0,\ldots,0)$, together with $(0,\ldots,0)$ with multiplicity $n$, a total of $(2n+1)(2n+2)/2 - 1$ weights. 

    \item \textbf{Antisymmetric rank$-k$} representation has weight decomposition given by all sums $\mu_1 + \ldots + \mu_k$ of distinct weights $\mu_i$ in the fundamental (vector) representation, i.e. of $n$-dimensional vectors $(\pm1,0,\ldots,0)$ and $(0,\ldots,0)$, for a total of $\binom{2n+1}{k}$ weights. 

    \item \textbf{Spinor} representation has weight decomposition given by all $n$-dimensional vectors of the form $\tfrac{1}{2}(\pm1,\ldots,\pm1)$, with dimension $2^n$. This representation is irreducible; therefore, in odd dimensions $D$ there are no Weyl spinors of definite chirality. 
\end{itemize}

Using the trace formula \eq{eq:trace-formula} and the weight decomposition of each representation we can compute all the necessary traces. 
Traces in the spinor representation are defined up to a sign, related to the choice of spin lift of $\mathbf{M}$ as discussed before; for any element $\D$ this sign is determined by the parameter $\sigma_\gamma$.
For example, taking the spinor representation with weight vectors $(\pm\frac12,\ldots,\pm\frac12)$ we find
\begin{equation}
    \Tr{\text{Spinor}}{\D} = (-1)^{2\sigma_\gamma} \prod_{i=1}^{n} 2\cos\frac{\theta_i}{2} \,.
    \label{eq:spinor-trace-weights}
\end{equation}

\hypertarget{TraceSpinor}{
\begin{mdframed}[style=function]
    \function{TraceSpinor[g,D,Weyl]} returns the trace of element \texttt{g} in the spinor representation of $SO(D-2)$, taking into account the spin lift of \texttt{g}; the parameter \texttt{Weyl} must be set to \texttt{True} for a Weyl spinor, and \texttt{False} otherwise (\texttt{Weyl = False} by default).
\end{mdframed}}

Finally, we defined the functions \function{TraceBosons} and \function{TraceFermions} that return the sum of the traces $\Tr{r}{\D}$ over all bosonic and fermionic representations, respectively. 
\hypertarget{TraceBosons}{
\begin{mdframed}[style=function]
    \function{TraceBosons[bosons,g,D]} returns the sum of all traces of element \texttt{g} under the bosonic representations of $SO(D-2)$ listed in \texttt{bosons}. 
\end{mdframed}}

\hypertarget{TraceFermions}{
\begin{mdframed}[style=function]
    \function{TraceFermions[fermions,g,D]} returns the sum of all traces of element \texttt{g} under the fermionic representations of $SO(D-2)$ listed in \texttt{fermions}.
\end{mdframed}}
The lists of \texttt{bosons} and \texttt{fermions} can be given in terms of \texttt{Associations} as described in the Section~\ref{sec:Casimir}. It is also possible to use the shorthand notation 
\begin{align*}
    \texttt{bosons} &= \texttt{\{graviton,p$_1$,p$_2$,\ldots\}} \\
    \texttt{fermions} &= \texttt{\{fermion$_1$,fermion$_2$,\ldots\}} \,.
\end{align*}
Set \texttt{graviton = 1} if the spectrum contains the graviton, and \texttt{graviton = 0} otherwise; for each $p$-form in the spectrum include the rank of the form, with $p=0$ for scalars; if the form is self-dual, input the corresponding rank as an imaginary number, $i\,p_1$. For fermions, set \texttt{fermion$_n$ = 3/2} for a gravitino (Rarita-Schwinger representation) and \texttt{fermion$_n$ = 1} for a spinor; to indicate that a fermion is in an irreducible Weyl representation (when $D$ is even) input the corresponding value as an imaginary number, \texttt{$i\,$fermion$_n$}. For example, 
\begin{itemize}
    \item[]\textbf{M-theory}: \texttt{spectrum} = \texttt{\{\{1,3\},\{3/2\}\};}
    \item[]\textbf{Type IIB}: \texttt{spectrum} = \texttt{\{\{1,0,0,2,2,4$\,i$\},\{3/2$\,i$,3/2$\,i$,1$\,i$,1$\,i$\}\}.}
\end{itemize}
The shorthand list notation does not support field-specific twisted boundary conditions. When this syntax is used, bosons are assigned periodic boundary conditions and fermions use the spin-structure vector \texttt{hspin}.

\section{Example: Type IIB on \texorpdfstring{$T^6/\Z_8$}{T6/Z8}}
\label{sec:example-TypeIIB}

As an example, consider Type IIB supergravity compactified on $T^6/\Z_8$ with the $\mathbb{Z}_8$ action generated by the affine transformation 
\begin{equation}
    \vec{z}\,\rightarrow\iota_{\mathbf{g}}(\vec{z})=\D[g]\,\vec{z}+\bvec[g] \,, 
    \quad\text{with}\quad 
    \D[g]\equiv\left(
        \begin{array}{*6{C{1.1em}}}
            0 & 0 & 0 & -1 & 0 & 0 \\
            1 & 0 & 0 & 0  & 0 & 0 \\
            0 & 1 & 0 & 0  & 0 & 0 \\
            0 & 0 & 1 & 0  & 0 & 0 \\ 
            0 & 0 & 0 & 0  & 1 & 0 \\
            0 & 0 & 0 & 0  & 0 & 1 \\
        \end{array}
        \right)\,,\quad \bvec[g]=\left(\begin{array}{c}0\\0\\0\\0\\0\\ \frac18\end{array}\right) \,.
    \label{Z8-rfmdef}
\end{equation}
We will choose the spin lift of $\D[g]$ such that $\sigma_\mathbf{g}=\frac12$. Using the function \function{GetGroup} we get the eight elements of $\Z_8$ generated by $\mathbf{g}$ (we show the first three elements).

{
    \flushleft\includegraphics[width=16cm]{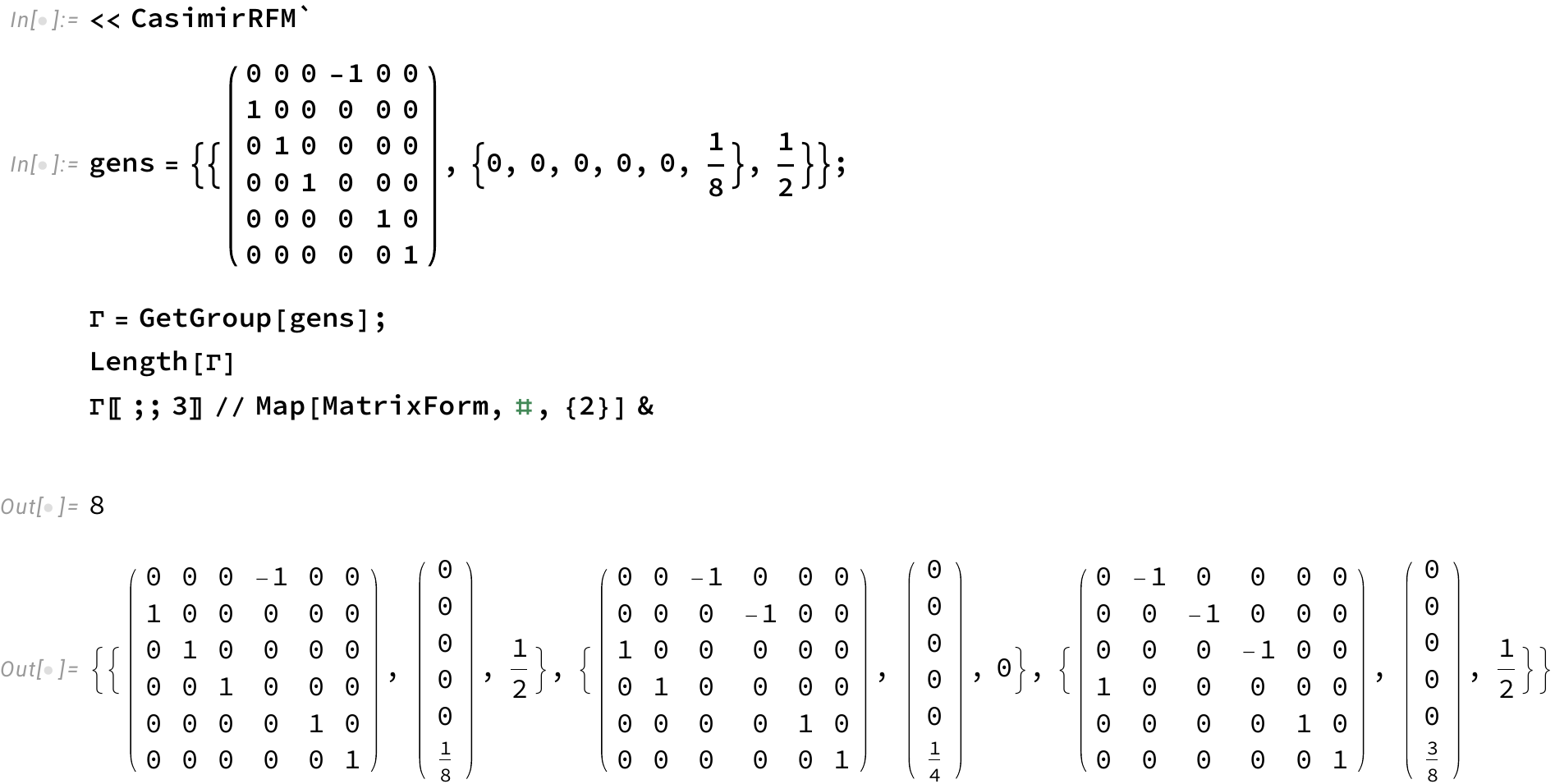}
}

\noindent
The most general metric invariant under this $\Z_8$ action contains 5 moduli and there are 4 distinct spin structures on $T^6$ compatible with the group action and spin lift. To get the most general form of the spin structure vector $\vec{h}$ we evaluate \texttt{SpinStructures[gens]["GeneralForm"]}, and to get a list of all allowed spin structures we evaluate \texttt{SpinStructures[gens]["Allowed"]}. We find that we must make the same choice for all of the first four directions, and that the last direction must be necessarily zero.

{
\flushleft\includegraphics[width=\textwidth]{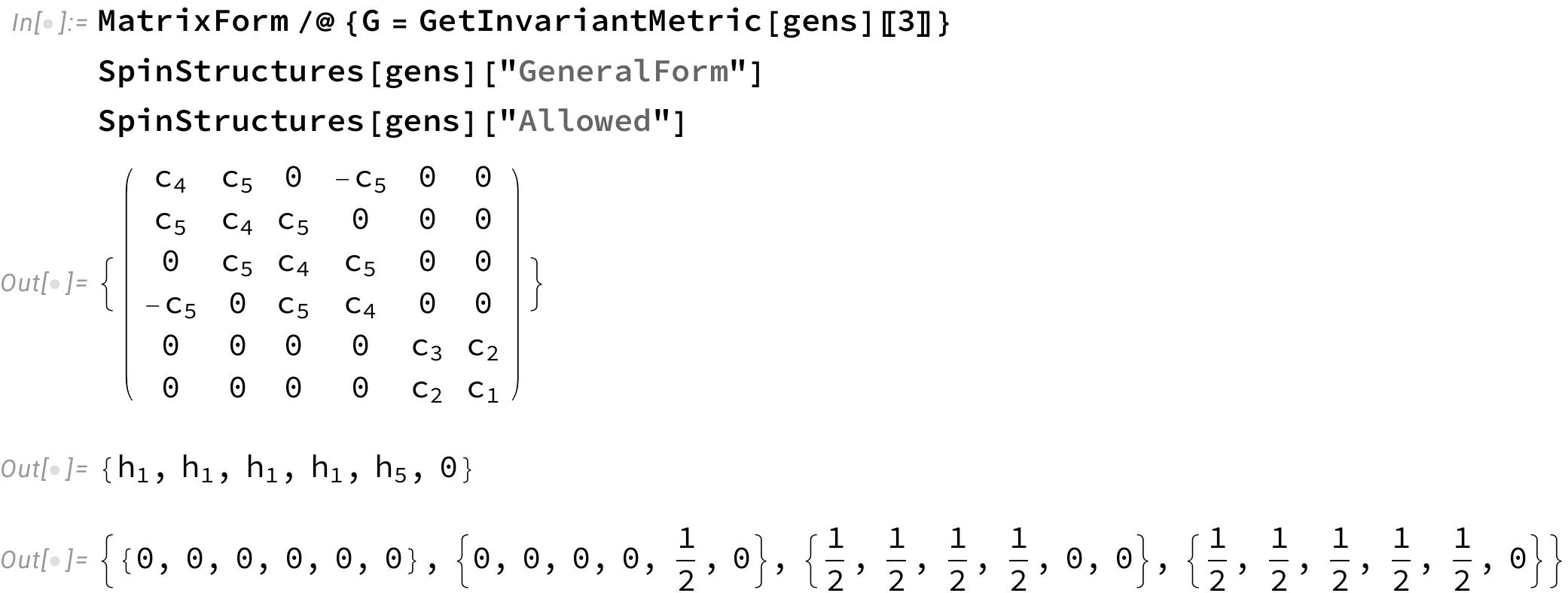}
}

The massless spectrum of Type IIB contains the graviton, two scalars, two 2-forms and one self-dual 4-form $(G,\phi,C_0,B_2,C_2,C_4)$, in the bosonic sector, and two Weyl spinors and two gravitinos $(\chi_\alpha^{\mu,1},\chi_\alpha^{\mu,2},\lambda_\alpha^{1},\lambda_\alpha^{2})$, in the fermionic sector. 
{\flushleft\includegraphics[width=0.8\textwidth]{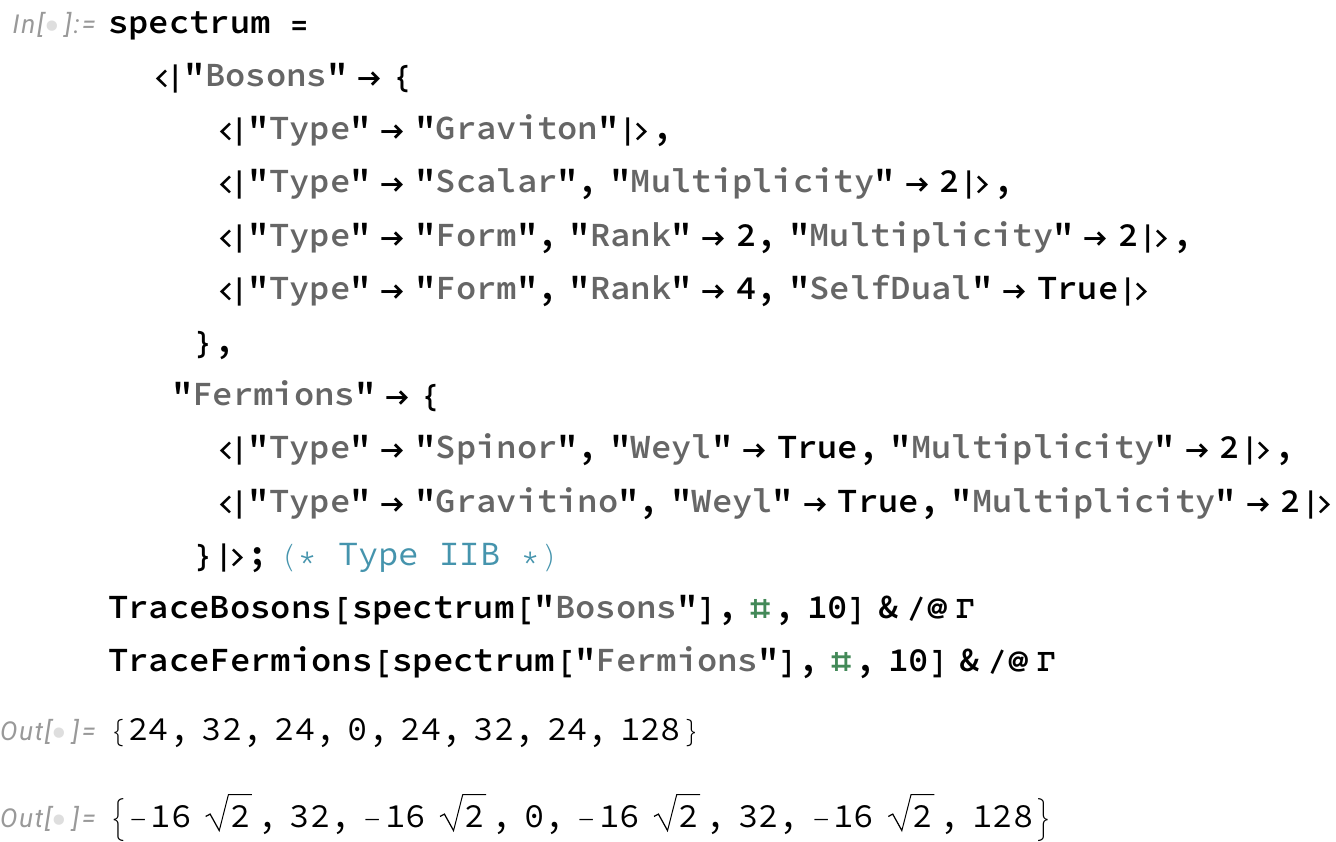}
\vspace{1em}
}

\noindent
For the identity element, both traces equal $128$, reproducing the matching numbers of physical bosonic and fermionic degrees of freedom in Type IIB supergravity.
The same results are obtained using the shorthand notation introduced in Section \ref{sec:traces}.
{ 
\flushleft\includegraphics[width=10cm]{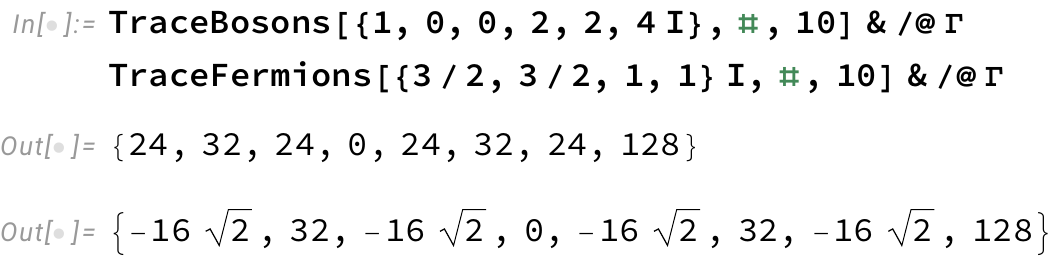}
\vspace{1em}
}

The Casimir potential is a function of the 5 moduli $\{c_1,c_2,c_3,c_4,c_5\}$. We can evaluate it---for example, at the point in moduli space that reduces the metric $G$ to the identity matrix---for two different choices of spin structure $\vec{h}$. The numerical values below use the default truncation target $\varepsilon = 10^{-4}$.
{
\flushleft\includegraphics[width=0.55\textwidth]{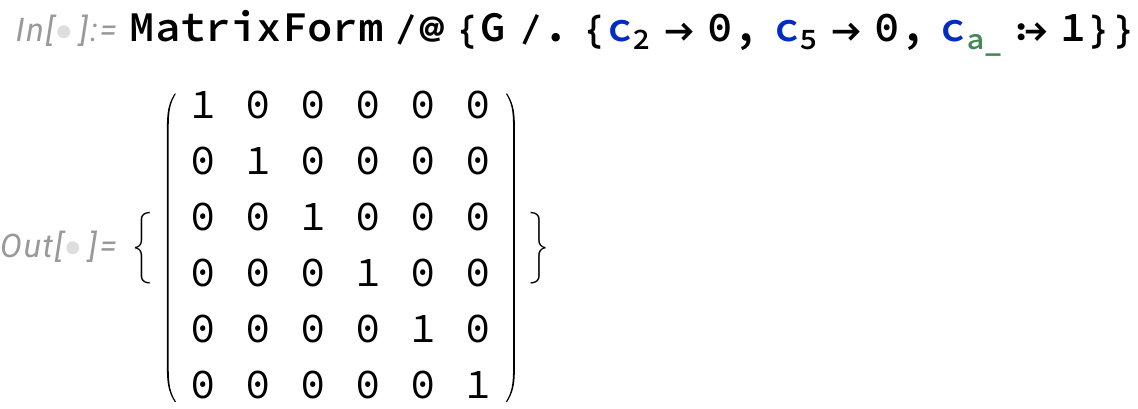}
}
{ 
\flushleft\includegraphics[width=\textwidth]{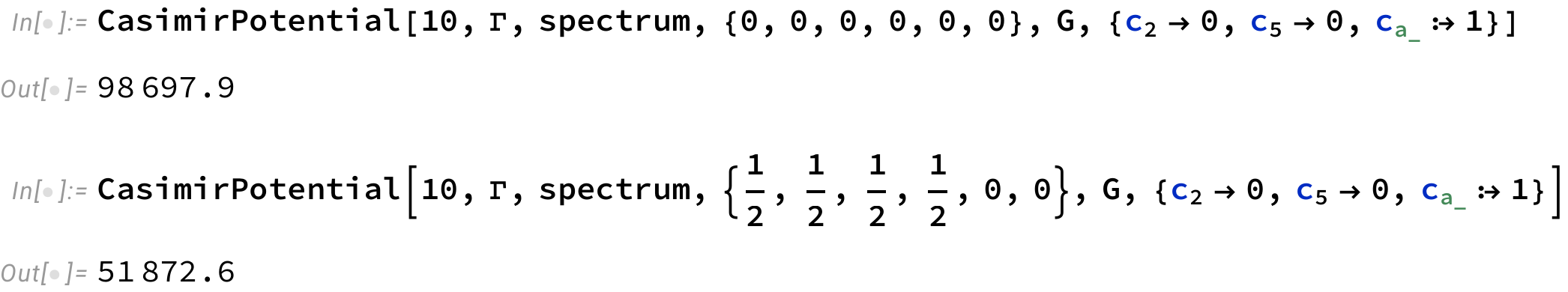}
}

By plotting the Casimir potential as a function of $c_2$, while keeping all other moduli fixed at their identity matrix values, we see that there is a critical point (maximum) along this direction at $c_2=0$, corresponding to a point in moduli space where the metric becomes the identity matrix. 
{ 
\flushleft\includegraphics[width=\textwidth]{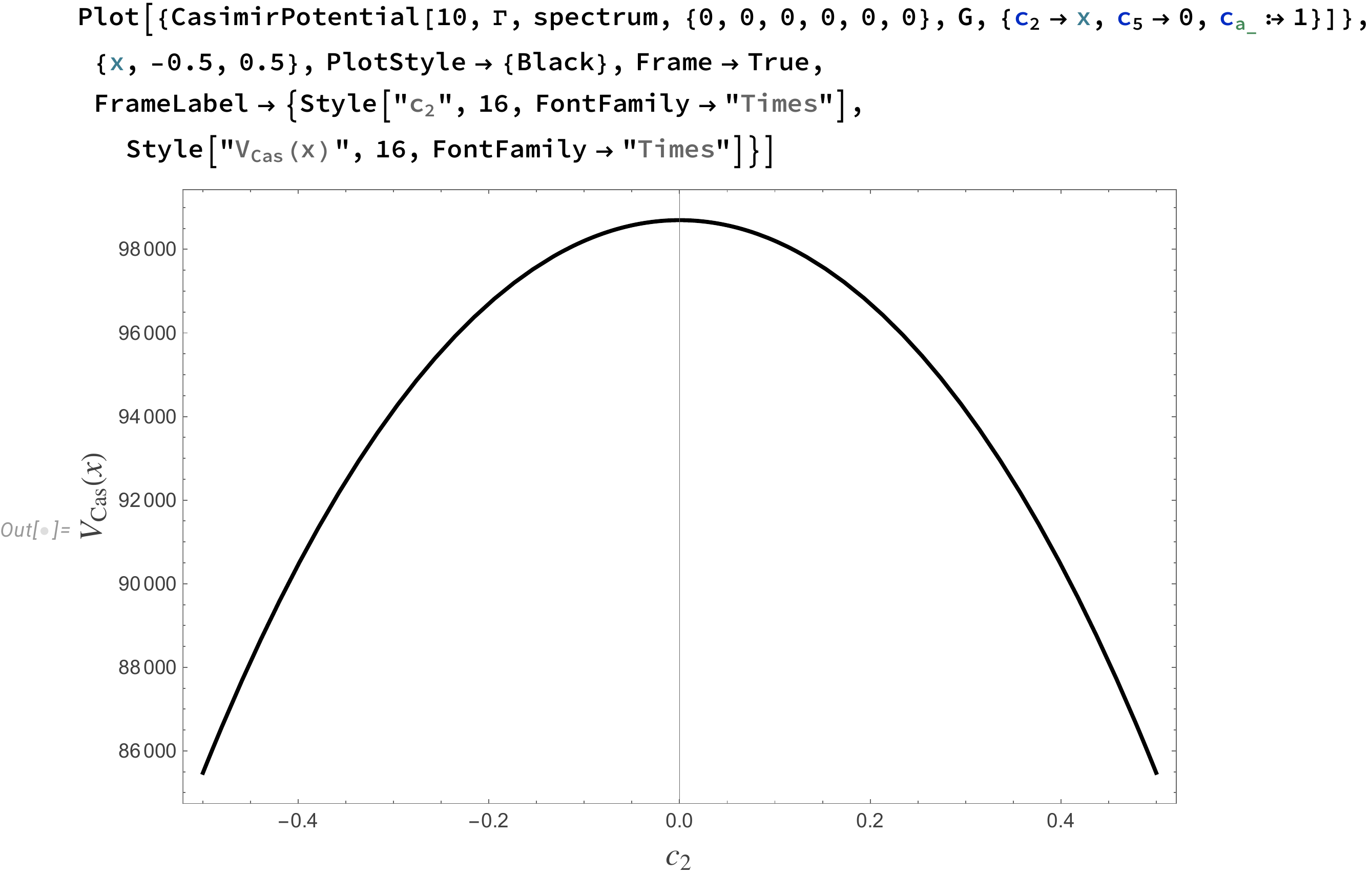}
}

We can visualise Casimir branes by plotting the Casimir energy density along a given plane. For example, there are two Casimir branes associated with the generator (and first element of $\Gamma$),
{
\flushleft\includegraphics[width=10cm]{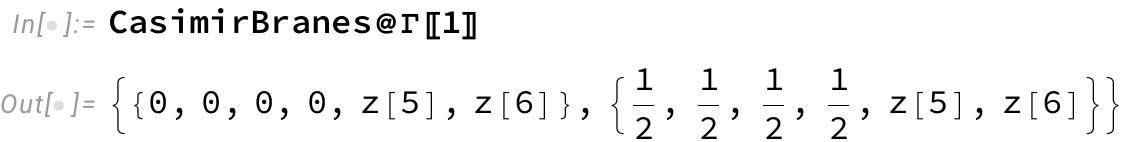}
\vspace{1em}
}

\noindent
The two branes are localised in the first four directions of $T^6$ and wrap the last two. In order to see both branes, we can change coordinates as
\begin{align}
    \begin{pmatrix}
        z_1\\z_2\\z_3\\z_4 
    \end{pmatrix}
    \to\begin{pmatrix}
        u_1\\u_2\\u_3\\u_4
    \end{pmatrix} = \frac14
    \begin{pmatrix}
        1 & 1 & 1 & 1 \\
        1 & -1 & 1 & -1 \\
        1 & 1 & -1 & -1 \\
        1 & -1 & -1 & 1
    \end{pmatrix}\cdot\begin{pmatrix}
        z_1\\z_2\\z_3\\z_4
    \end{pmatrix} \,,
\end{align}
so that the two Casimir branes are located at $\{0,0,0,0\}$ and $\{\frac12,0,0,0\}$ and we can plot the Casimir energy density along the $u_1$--$u_2$ plane, with $u_3=u_4=0$.
{\vspace{1em}
\flushleft\includegraphics[width=10cm]{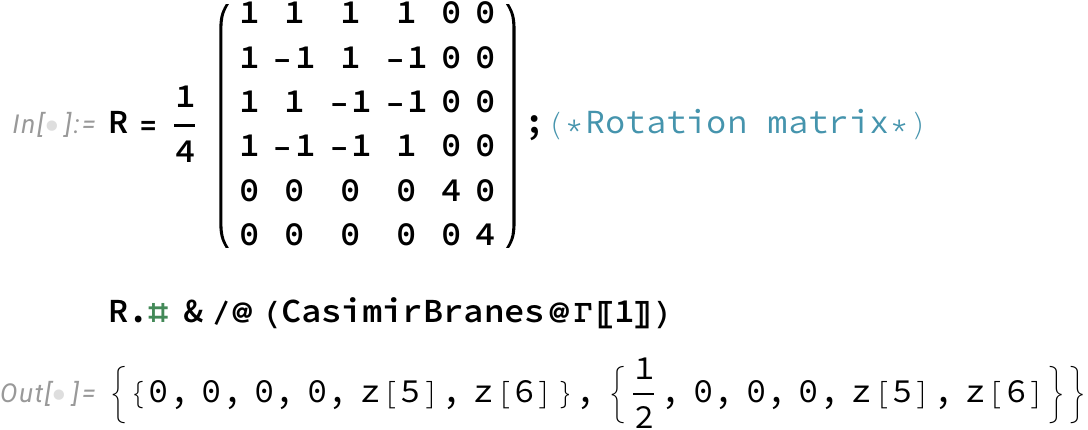}
\vspace{1em}
}

\noindent
This is also a case in which parallelising the computation is particularly useful. We can first generate a \texttt{ParallelTable} with 
\begin{enumerate}
    \item the Casimir energy density from element $\Gamma[[1]]$;
    \item the boson contribution to the Casimir energy density from element $\Gamma[[1]]$;
    \item the fermion contribution to the Casimir energy density from element $\Gamma[[1]]$;
\end{enumerate}
and then plot them using \texttt{ListPlot3D}. This also has the advantage that we can style the plot without having to evaluate the functions every time. 

{
\flushleft\includegraphics[width=0.5\textwidth]{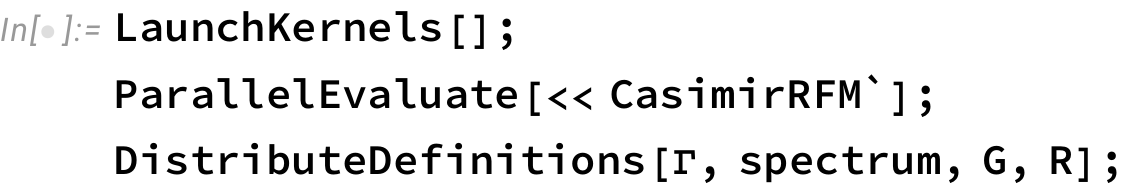}
}

{
\flushleft\includegraphics[width=\textwidth]{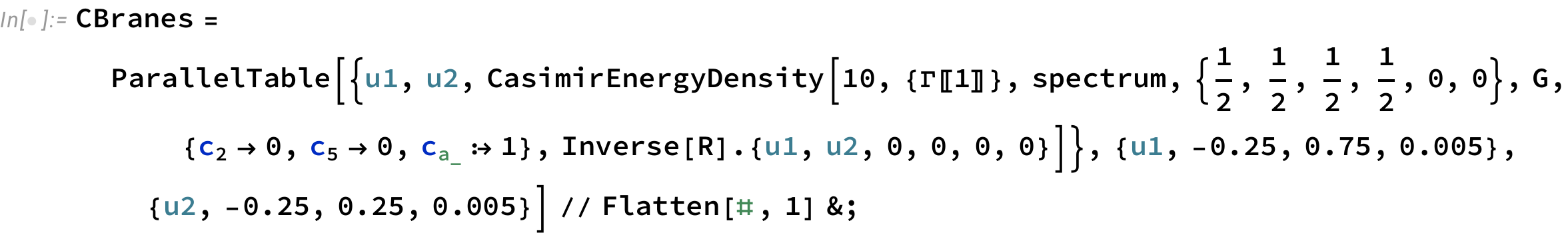}
\flushleft\includegraphics[width=\textwidth]{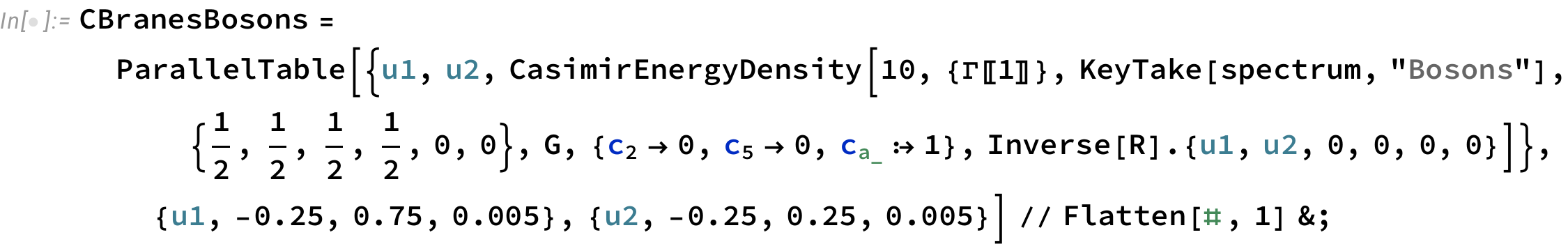}
}
{
\flushleft\includegraphics[width=\textwidth]{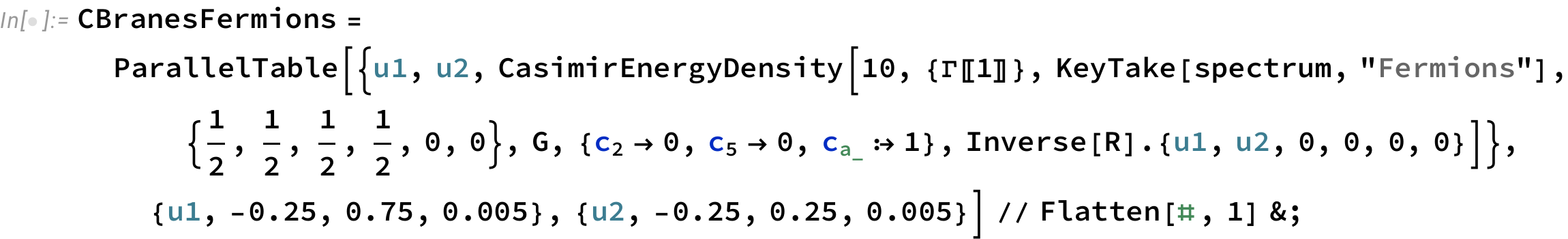}
\flushleft\includegraphics[width=\textwidth]{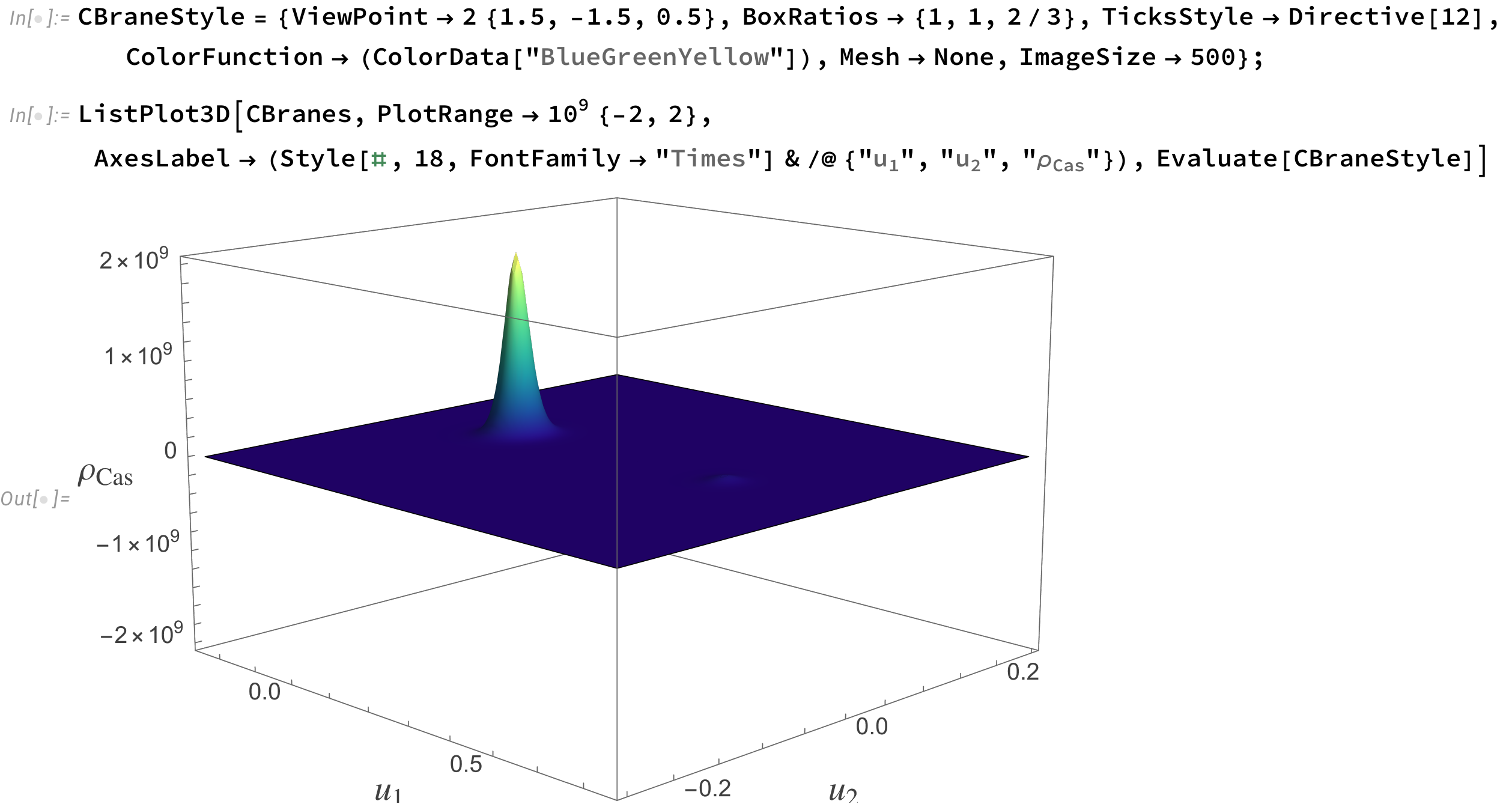}
}

\noindent
By plotting the boson and fermion contributions separately we see the cancellation between them at $\{\frac12,0,0,0\}$, which explains why this Casimir brane is not visible in the plot of the full Casimir energy density. 
{\vspace{1em}
\flushleft\includegraphics[width=\textwidth]{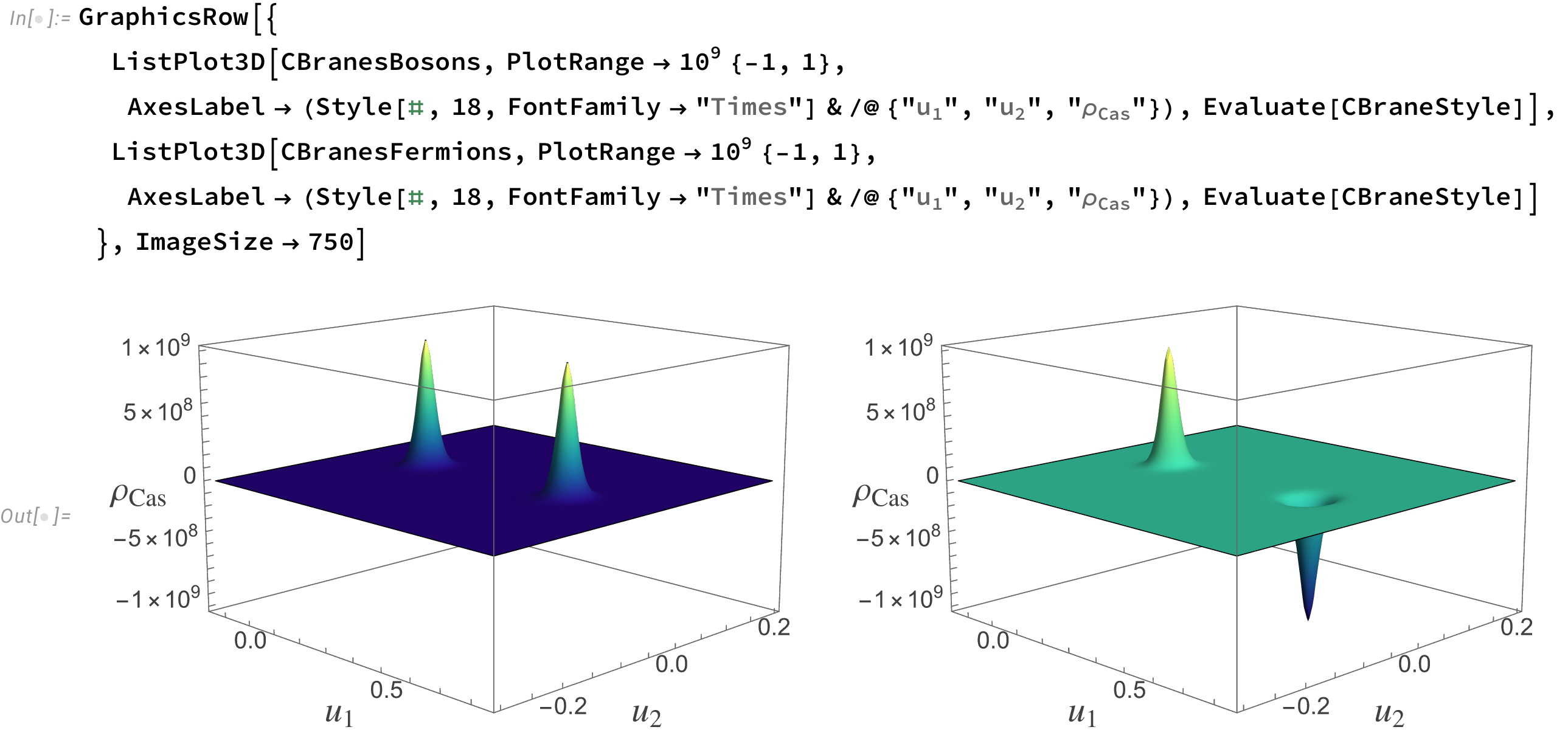}}

The notebook \texttt{CasimirRFM-TypeIIB-Example.nb} used for this example is included as suplementary material and can also be found in the package \href{https://github.com/bruno-valeixo-bento/CasimirRFM}{repository}, together with the notebook \texttt{CasimirRFM-Template.nb}, providing descriptions and examples for every function.

\vskip 5em
\noindent\textbf{Acknowledgements:} 
We thank Miguel Montero for his collaboration on the development of the methods implemented in this package and early implementation of some of the functions; Miquel Aparici for helpful discussions and for testing progressive versions of the package; and Michelangelo Tartaglia for helpful discussions on spin structures and spin lifts, much needed for the conclusion of the package, and feedback on the manuscript. 
We gratefully acknowledge the support of an Atraccion del Talento Fellowship 2022-T1/TIC-23956 from Comunidad de Madrid, which supported this project in its early stages, as well as the Spanish State Research Agency (Agencia Estatal de Investigacion) through the grants Europa Excelencia EUR2024-153547, IFT Centro de Excelencia Severo Ochoa CEX2020-001007-S, PID2021-123017NB-I00 and PID2024-156043NB-I00, funded by MCIN/AEI/10.13039/501100011033, and ERDF, EU. 

\bibliographystyle{utphys}
\bibliography{refs}

\end{document}